\documentclass[a4paper,12pt]{article}
\usepackage{jcappub} 

\title{\boldmath Comprehensive Study of $k$-essence Model: Dynamical System Analysis and Observational Constraints from Latest Type Ia Supernova and BAO Observations}

	\author[1]{Saddam Hussain,}
	\emailAdd{mdsaddamh6@gmail.com}
	
	\author[2]{Sarath Nelleri,}
	\emailAdd{sarathn@iitk.ac.in}
	
	\author[3]{Kaushik Bhattacharya}%
	\emailAdd{kaushikb@iitk.ac.in}

	\affiliation[1,2,3]{%
		Department of Physics, Indian Institute of Technology Kanpur, Kalyanpur 208016, India
	}%
	
	\abstract{We constrain the parameters of the $k$-essence scalar field model with inverse square and exponential potentials using data sets including Pantheon+SHOES and the Dark Energy Survey (DES) of Type Ia supernovae, Baryon Acoustic Oscillation (BAO) data from SDSS and DESI surveys, and direct measurements of the Hubble parameter and redshift obtained from the differential age method (CC). We also provide a brief perspective on the dynamical evolution of both models and derive stability constraints on the model parameters, which are then used to set appropriate priors. We adopt a Bayesian inference procedure to estimate the model parameters that best fit the data. A comprehensive analysis in light of observational data shows that the $k$-essence model fits well across all data combinations. However, according to the BIC criterion, the $\Lambda$CDM model provides a slightly better fit compared to the $k$-essence model.}

	\usepackage{tabularx}
	\usepackage{graphicx}
	\usepackage{bm}
	\usepackage{newtxmath,newtxtext,amsfonts}
	\usepackage{xcolor}
	\usepackage{hyperref}
	\usepackage{multirow}
	\usepackage{float}
	\usepackage[caption=false]{subfig}
	\hypersetup{
		colorlinks=true,
		linkcolor=blue,
		filecolor=violet,     
		urlcolor=blue,
		citecolor=red
	}

	\newcommand{\ti}[1]{\ensuremath{ \tilde{#1}}}

	\newcommand{\es}{\ensuremath{\omega_{\rm eff}}}

	\newcommand{\om}{\ensuremath{\Omega_{m}}}
	
	\usepackage{makecell}

\begin{document}
		\maketitle
		\flushbottom
		
		\section{Introduction}

		Initially, the observation on Type Ia supernovae \cite{SupernovaCosmologyProject:1998vns,SupernovaSearchTeam:1998fmf}, subsequent independent observations from the Cosmic Microwave Background (CMB), gravitational lensing, large-scale structure, and Baryonic Acoustic Oscillations (BAO) \cite{WMAP:2003elm,Sherwin:2011gv,Wright:2007vr,DES:2016qvw,DES:2021esc,SDSS:2005xqv}, indicated the existence of an exotic energy component dubbed dark energy, responsible for the accelerating expansion of the universe. The simplest and widely accepted model that accounts for this observation is $\Lambda$CDM  where $\Lambda$ represents the cosmological constant and CDM refers to cold dark matter.
		However, it poses challenges such as fine tuning problem, cosmic coincidence, Hubble tension among others. The fine-tuning problem refers to a profound disparity of approximately 120 orders of magnitude between the observed value ($\sim 10^{-47}\textrm{GeV}^4$) and the theoretically predicted value ($\sim 10^{74}\textrm{GeV}^4$) of the cosmological constant ($\Lambda$), assuming it represents the vacuum energy density. The cosmic coincidence problem puzzles over why the densities of dark energy and dark matter appear to be nearly equal in the present epoch \cite{Copeland:2006wr,Weinberg:1988cp,Rugh:2000ji,Padmanabhan:2002ji,Carroll:1991mt,Bengochea:2019daa,Kohri:2016lsj,Lopez-Corredoira:2017rqn}. The tension problem in cosmology arises from the statistically significant ($4\sigma$ to $6\sigma$) discrepancy between the cosmological parameters predicted by early-time data assuming the concordance $\Lambda$CDM model and a wide range of model-independent local measurements of distances and redshifts. For instance, the Hubble constant ($H_0$) estimated from CMB data by the Planck2018 collaboration is $67.4\pm0.5$ $\textrm{km} \ \textrm{s}^{-1}\textrm{Mpc}^{-1}$ \cite{aghanim2020planck}, whereas direct observations of Cepheids by the SH0ES team report $73.2\pm 1.3$ $\textrm{km} \ \textrm{s}^{-1}\textrm{Mpc}^{-1}$ \cite{riess2021cosmic}, indicating a tension of approximately $\sim 4.2\sigma$. 
		
		Over the past two decades, researchers have explored various alternative scenarios that can mimic the dynamics of $\Lambda$CDM. One of the most popular and heavily investigated mechanisms to tackle this problem is via the introduction of a scalar field which acts as the dark energy constituent. Inspired by particle physics, a quintessence-type field with a potential term can generate sufficient acceleration in the presence of dark matter \cite{Peebles:2002gy,Nishioka:1992sg,Ferreira:1997hj,Zlatev:1998tr,Copeland:1997et,Hebecker:2000au,Hebecker:2000zb,Roy:2022fif,Hussain:2023kwk,Das:2023rat}. Although the quintessence field minimally coupled with dark matter fluid can mimic the dynamical aspect of $\Lambda$CDM, it also suffers from fine-tuning of the potential parameter. Later, some string-inspired models, namely $k$-essence fields with non-linear kinetic terms, have been widely studied. It is claimed  that these theories can address some of the shortcomings of the above models. Due to the associated non-linear kinetic terms, these fields have been termed non-canonical fields \cite{Armendariz-Picon:2000nqq,Armendariz-Picon:2000ulo,Chiba:1999ka,Fang:2014qga,Armendariz-Picon:1999hyi,Armendariz-Picon:2005oog,Arkani-Hamed:2003pdi,Scherrer:2004au,Chatterjee:2021ijw,Hussain:2022osn,Bhattacharya:2022wzu,Hussain:2022dhp}. {Despite its success in explaining late-time cosmic acceleration, only a small number of attempts have been made to test the $k$-essence model against observational data to see how well they fit with late-time cosmological observations.}

		{Testing the $k$-essence model against observational data is crucial to determine how well it aligns with the observations. To perform this analysis, we numerically solve the field and Friedmann equations, thereby obtaining the Hubble parameter as a function of redshift. Setting initial conditions and priors for the model parameters involves conducting a dynamical system analysis. Stability techniques within dynamical systems are invaluable tools for examining complex models, as they facilitate the identification of physically viable solutions by analyzing the existence and nature of critical points, thus constraining the model parameters \cite{Coley:2003mj, Rendall:2001it, Boehmer:2011tp, Dutta:2017wfd, blackmore2011nonlinear, Bouhmadi-Lopez:2016dzw, elias2006critical, Bahamonde:2017ize,Amendola_Tsujikawa_2010,Copeland:2023zqz,Barreiro:1998aj,Dutta:2016bbs,Geng:2015fla,Yongjian2021,ELIAS2006305,Ng:2001hs,Chakraborty:2024nmg,Yang:2010vv,Chen:2022vmi,Chen:2008ft}. These critical points correspond to the various phases of the universe. Some of the critical points are practically useful as they may have a connection to the late-time acceleration of the universe.
			
			To assess the stability of the critical points, we establish a system of first-order autonomous differential equations. By linearizing these autonomous equations, we can construct a Jacobian matrix and calculate the corresponding eigenvalues. If any real part of the eigenvalues is zero, the linearization method may not be applicable for determining stability. In such cases, more advanced mathematical approaches, such as the center manifold theorem, become necessary to accurately evaluate the stability.
		}
		
		{In this study, we analyze the $k$-essence scalar field with the Lagrangian as \(\mathcal{L}_{\phi} = V(\phi) (-X + X^2)\), where \(X = -\frac{1}{2}g^{\mu\nu}\nabla_{\mu} \phi \nabla_{\nu}\phi\), with two types of potentials: (i) inverse square potential \(V(\phi) \propto \frac{1}{\phi^2}\) and (ii) exponential potential \(V(\phi) \propto e^{C\phi}\) ($C$ is a dimensional constant). We test these models using observational datasets including Cosmic Chronometer (CC), Baryonic Acoustic Oscillation (BAO), Dark Energy Survey Supernova 5YR (DES-SN), and SH0ES-calibrated Pantheon samples (Pantheon+SH0ES) of Type Ia supernovae. While extensive research has been conducted on the $k$-essence field with the inverse square potential, there is limited literature addressing the exponential potential within the Lagrangian framework specified above. Moreover, studies on pure kinetic $k$-essence fields are sparse \cite{Yang:2009zzl, Dinda:2023mad}. Our analysis indicates that the exponential potential model can exhibit a late-time cosmic accelerating phase, preceded by a matter dominated phase that persists for a brief period.
			
			The data analysis reveals that both models are consistent with the observational datasets, presenting a compelling alternative to the $\Lambda$CDM model. The Akaike Information Criterion (AIC) and Bayesian Information Criterion (BIC) both strongly support the current models over $\Lambda$CDM when considering the combined datasets.  }
		
		The paper is organized as follows: In Sec. \ref{sec:3}, we analyze the $k$-essence model with inverse square and exponential type potentials from a dynamical system perspective and present the numerical evolution of the cosmological parameters. The data analysis of the $k$-essence models is discussed in Sec. \ref{sec:data_analysis}. Finally, we present our conclusions in Sec. \ref{sec:conclusion}.
		
		\section{K-essence field}
		\label{sec:3}
		The action for the \(k\)-essence field, minimally coupled with the dark matter fluid, is given by:
		\begin{equation}
			S = \int d^4 x \sqrt{-g} \bigg(\frac{R}{2\kappa^2} -  \mathcal{L}_{\phi}(X,\phi) +  \mathcal{L}_{m}\bigg)   ,
		\end{equation}
		{where the $k$-essence Lagrangian $\mathcal{L}_{\phi}$ is related to the pressure of the field, and $\mathcal{L}_{m}$ denotes the Lagrangian for the perfect fluid. In our system of units $\kappa^2=8\pi G$ where $G$ is Newton's universal gravitational constant.} The pressure of the \(k\)-essence field is assumed to have a particular form as follows:
		\begin{equation}
			P_{\phi} = -\mathcal{L}_{\phi} = V(\phi) F(X).
		\end{equation}
		Here, \(V(\phi)\) represents the potential of the field, and \(F(X)\) depends on the kinetic term of the field, where \(X = -\frac{1}{2} g^{\mu\nu} \partial_{\mu}\phi \partial_{\nu} \phi\). The potential extensively studied corresponding to this particular form of the Lagrangian, which exhibits scaling behavior in the past epoch and tracking behavior in the current epoch, is the inverse square type of potential \cite{Copeland:2006wr,Armendariz-Picon:1999hyi,Jorge:2007zz}. {However, in this paper our study is not limited to this particular type of potential. We will expand our study to include two types of potentials, namely the inverse square and exponential type potentials, referred to in the text as Model I and Model II. The forms of the potentials are given by: 
			\begin{equation}
				\text{Model I:} \quad V(\phi) = \frac{\delta^2}{\kappa^2 \phi^{2}}, \quad \text{Model II:} \quad V(\phi) = V_0 e^{\beta H_0 \phi}. 
			\end{equation}
			Here, the parameters $\delta$ and $\beta$ are dimensionless constants, whereas \(H_0\) refers to a constant whose dimension is the same as the Hubble parameter. The other parameter \(V_0\) is a dimensional constant with mass dimension four. We have considered the exponential potential because this type of potential is usually studied with quintessential types of scalar fields \cite{Copeland:1997et}. However very few studies, involving statistical analysis, have been conducted with the $k$-essence field \cite{Rendall:2005fv,Chimento:2005ua,Orjuela-Quintana:2021zoe}, particularly with the Lagrangian form $\mathcal{L}_{\phi} = -V(\phi) F(X)$. Hence, we will carry out a detailed study of the dynamics of the $k$-essence field with both potentials. To study the following scenario, we will take the particular form of the kinetic function as given by: }
		\begin{equation}
			F(X) = - X + X^2.
		\end{equation}
		Upon varying the action of the \(k\)-essence with respect to \(g^{\mu\nu}\), the stress tensor of the field can be obtained as:
		\begin{eqnarray}
			T_{\mu \nu}^{(\phi)} = - {\mathcal L}_{,X}\,(\partial_\mu \phi)(\partial_\nu \phi)-g_{\mu \nu}\,{\mathcal L}\,.
			\label{tphi}  
		\end{eqnarray}
		Corresponding to this, the energy density and pressure of the \(k\)-essence field can be obtained as follows: 
		\begin{equation}\label{}
			\rho_{\phi} = \mathcal{L} - 2X\mathcal{L}_{,X} \quad \text{and} \quad P_{\phi} = - \mathcal{L}\,.
		\end{equation}
		As the field and fluid components are minimally coupled with gravity, the corresponding individual energy-momentum tensors are conserved: 
		\begin{equation}
			\begin{split}
				\dot{\rho}_{m} + 3 H (\rho_{m} + P_{m}) &= 0,\\
				\dot{\rho}_{\phi} + 3 H (\rho_{\phi} + P_{\phi} ) &= 0 .
			\end{split}
		\end{equation}
		{Here, $\rho_{m}$ and $P_{m}$ denote the background matter fluid energy density and pressure, respectively. In this paper, we choose the background fluid to be pressureless, i.e., $P_{m} = \omega_{m} \rho_{m} = 0$.} Upon varying the action with respect to \(\phi\), the field equation of motion becomes:  
		\begin{equation}
			\label{motion1}
			(F_{,X} + 2X F_{,XX})\ddot \phi + 3H F_{,X} \dot \phi + (2XF_{,X} - F)\frac{V_{,\phi}}{V} = 0,
		\end{equation}
		where the subscript associated with the comma denotes the partial derivative with respect to that quantity. {In the spatially flat Friedmann-Lemaître-Robertson-Walker (FLRW) metric, \(ds^2 = -dt^2 + a(t)^2 d\vec{x}^2\), where \(a\) denotes the scale factor, the Friedmann equations become:}
		\begin{eqnarray}
			3H^2 &=& \kappa^2(\rho_{m} + \rho_{\phi}),\label{kes_frd1}\\
			2\dot{H} + 3H^2 &=& -\kappa^2 (\rho_{m} + \rho_{\phi}).
		\end{eqnarray}
		In the subsequent section, we translate these equations into autonomous equations and discuss  the stability of the system.
		
		\subsection{Stability of the system in the presence of Inverse square potential}
		
		After evaluating the background dynamical equations in the aforementioned section, we will analyze the stability of the system {corresponding to the inverse square type of potential} by defining a set of dimensionless variables as:
		\begin{equation}
			x^2= \frac{\kappa^2 V \dot{\phi}^2}{6 H^2}, \quad y^2 = \frac{\kappa^2 V \dot{\phi}^4}{4 H^2}, \quad \Omega_{m} = \frac{\kappa^2 \rho_{m}}{3 H^2}, \quad \lambda = \frac{- V_{,\phi}}{\kappa V^{3/2}}, \quad  \Omega_{\phi} = \frac{\kappa^2 \rho_{\phi}}{3 H^{2}}.
		\end{equation}
		{Although a large section of the literature has been published discussing the dynamics corresponding to the inverse square potential, we are explicitly showing it here for the sake of consistency. To analyze the dynamics, we choose a different dynamical setup to comprehend the dynamics of the phase space. While the dynamical variables are defined differently, the results we obtained match those of previous studies.} We choose the variables corresponding to the field sector such that the Hubble equation takes a similar form to the quintessence field as presented in Ref.~ \cite{Copeland:2006wr}. The primary variables associated with the \(k\)-essence field are \((x, y > 0)\), and are assumed to be positive. {The variables $(\Omega_{\phi}, \Omega_{m})$ are secondary variables as they can be expressed in terms of primary variables. The other variable, $\lambda$, is called the slope of the potential, which in this case depends on $\delta$ and becomes a constant. It should be noted that the dynamical variables defined here will not be valid for all types of potential. For instance, the dynamical variables chosen for the exponential potential in the next section in Eq.~\eqref{dyn_variab_exponential}, although using similar notations, have definitions that are quite different from those used here.} Following the Hubble equation Eq.~\eqref{kes_frd1}, the primary variables are \((x,y, \Omega_{m})\) constrained as:
		\begin{equation}
			0 \le \Omega_{m} = 1- \Omega_{\phi} \le 1, \quad 0 \le \Omega_{\phi} \le 1.
			\label{min_constraint}
		\end{equation}
		Here, the field density parameter, \(\Omega_{\phi}\), and effective equation of state, \(\es\), in terms of dynamical variables become:
		\begin{equation}
			\Omega_{\phi} = -x^2 + y^2, \quad \es = \frac{P_{m} + P_{\phi}}{\rho_{m} + \rho_{\phi}} = \frac{y^2}{3}-x^2. 
		\end{equation}
		Similarly, the second Friedmann equation can be expressed in terms of dynamical variables as:
		\begin{equation}
			\frac{\dot{H}}{H^2} = \frac{3}{2} \left(x^2-\frac{y^2}{3}-1\right).
		\end{equation}
		To understand the system's dynamics, we take the primary variables and set up the autonomous equations as follows: 
		\begin{equation}
			\label{eq:2}
			\begin{split}
				x'  &= -\frac{\dot{H}}{H^2} x-\frac{1}{2} \sqrt{6} \lambda x^2+\frac{-\frac{1}{2} \sqrt{6} \lambda  x^2-\frac{2 y^2}{x}+3 x+\frac{1}{2} \sqrt{6} \lambda  y^2}{\frac{2 y^2}{x^2}-1},\\
				y' &=-\frac{\dot{H}}{H^2} y+\frac{-\frac{4 y^3}{x^2}+\frac{6 \lambda  y^3}{\sqrt{6} x}-\sqrt{6} \lambda  x y+6 y}{\frac{2 y^2}{x^2}-1}-\frac{1}{2} \sqrt{6} \lambda  x y\ .
			\end{split}
		\end{equation}
		Here, the time derivative of the dynamical variables is closed and requires no extra variables, as a result, the phase space of the system becomes \(2D\). The autonomous equations \((x',y')\) present the invariant submanifold \((x = 0, y = 0)\), hence the trajectories that originate in \(x, y>0\) cannot go out from this region. For the inverse square type potential, the derivative of the potential yields:
		\begin{equation}
			\lambda = 2/\delta.
		\end{equation}
		As \(\lambda\) becomes a constant quantity, throughout the analysis, we treat \(\lambda\) as a model parameter instead of \(\delta\). Now, we will obtain the critical points associated with the autonomous equations and summarize them in Tab. [\ref{tab:cric_kes_minimal}].
		\begin{table}[t]
			\centering
			\tiny
			\begin{tabular}{cccccc}
				\hline
				Points & $(x,y)$ & $\Omega_{\phi}$ & $\om$ & $\es$ & Eigenvalues\\
				\hline
				$P_{1}$ & $(0,1)$ & $1$ &$0$ & $1/3$ &\((1,1)\) \\
				\hline 
				$P_{2}$ & $\left(\frac{\sqrt{\frac{3}{2}}}{\lambda },\frac{3}{\sqrt{2} \lambda }\right)$ & $\frac{3}{\lambda ^2}$ & $1-\frac{3}{\lambda ^2}$ & 0 & $\left(-\frac{3 \sqrt{\frac{3}{5}} \sqrt{-\lambda ^2 \left(\lambda ^2-8\right)}}{4 \lambda ^2}-\frac{3}{4},\frac{3}{20} \left(\frac{\sqrt{15} \sqrt{-\lambda ^2 \left(\lambda ^2-8\right)}}{\lambda ^2}-5\right)\right)$ \\
				\hline
				$P_{3}$&$\left(-\frac{\sqrt{3 \lambda ^2+16}+\sqrt{3} \lambda }{2 \sqrt{2}},\frac{1}{2} \sqrt{3 \lambda ^2+\sqrt{9 \lambda ^2+48} \lambda +12}\right)$ & $1$ & 0 & $\frac{1}{6} \left(-3 \lambda ^2-\sqrt{9 \lambda ^2+48} \lambda -6\right)$ & Fig. [\ref{fig:p3_stab_kes}] \\
				\hline 
				$P_{4}$ & $\left(\frac{\sqrt{3 \lambda ^2+16}-\sqrt{3} \lambda }{2 \sqrt{2}},\frac{1}{2} \sqrt{3 \lambda ^2-\sqrt{9 \lambda ^2+48} \lambda +12}\right)$ & 1 & 0 & $\frac{1}{6} \left(-3 \lambda ^2+\sqrt{9 \lambda ^2+48} \lambda -6\right)$ & Fig. [\ref{fig:p4_stab_kes}]\\
				\hline
				$P_{5}$ & $(0,0)$ & $0$ & $1$ & 0 & $\left(-3/2, -9/2\right)$ \\
				\hline
				\hline
			\end{tabular}
			\caption{The nature of the critical points of the $k$-essence field. }
			\label{tab:cric_kes_minimal}
		\end{table}
		The system produces four critical points, and we have considered only those points for which \(y>0\). {As the matter sector is assumed to be made up of dark matter, our assumption implies that the fluid representing dark matter is pressure-less $\omega_{m} = 0$. We use linearization techniques to determine the stability of the critical points by evaluating the Jacobian matrix defined as follows:
			\begin{equation}
				J_{ij} = \frac{\partial x_{i}'}{\partial x_{j}}\bigg|_{\bm{x}_*},
			\end{equation}
			where \(x_i = \{x,y,...\}\) for \(i = 1,2,...\) and $\bm{x}_*$ represent the set of coordinates specifying a critical point. Corresponding to this autonomous system, the Jacobian matrix becomes \(n \times n\) dimensional. To assess stability, we determine the eigenvalues and based on the signature of the real parts of eigenvalues, the point can be categorized as stable, unstable, or saddle. If all the real parts of the eigenvalues are negative (positive), the corresponding point becomes asymptotically stable (unstable). However, for alternating signs, the point is labeled as a saddle. The detailed description of the critical points and corresponding stability is discussed as follows:
		}

		\begin{itemize}
			\item \textbf{Point \(P_{1}\):} At this point, the field energy density parameter dominates over the fluid energy density parameter and produces an effective equation of state, \(\es = 1/3\), showing radiation-type characteristics. This signifies that the \(k\)-essence field can mimic radiation-type behavior in an early epoch of the universe. The eigenvalues at this point yield positive values, indicating unstable behavior.
			
			\item \textbf{Point \(P_{2}\):} At this point, the coordinates become model parameter dependent. Both the field and fluid density parameter are non-zero and dependent on \(\lambda\). The point can be physically viable for \(\lambda^2 \ge 3\). The effective equation of state at this point vanishes, showing pressureless matter fluid-type characteristics. Based on the eigenvalues, the point becomes an attractor point for \(\lambda \ge \sqrt{3}\). However, as the point becomes an attractor point in some range of \(\lambda\), and produces non-accelerating type behavior, it cannot be considered as a viable physical point that describes the late-time phase of the universe.
			
			\item \textbf{Point \(P_{3}\):} The coordinates are \(\lambda\) dependent, and the field density parameter dominates over the fluid density. The effective EoS, \(\es\) is \(\lambda\) dependent and produces an accelerating solution, i.e., \(-1 \leq \es \leq -1/3\), for \(-0.82<\lambda <0\) and a phantom solution, \(-1.2 \leq \es <-1\), for \(0<\lambda <0.162\). In this range, the real part of the eigenvalues are negative, as shown in Fig. [\ref{fig:p3_stab_kes}], hence the point becomes stable and shows the late-time behavior of the universe.
			
			\begin{figure}[t]
				\centering
				\subfloat[\label{fig:p3_stab_kes}]{\includegraphics{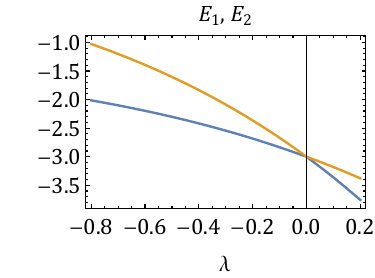}}
				\hspace{0.3cm}
				\subfloat[\label{fig:p4_stab_kes}]{\includegraphics{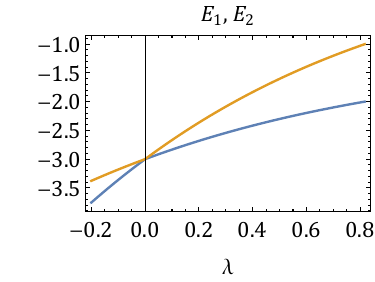}}
				\caption{The evolution of eigenvalues at points $P_{3,4}$ against $\lambda$ shown in Figs. [\ref{fig:p3_stab_kes}, \ref{fig:p4_stab_kes}] respectively. }
			\end{figure}
			
			\item \textbf{Point \(P_{4}\):} Similar to point \(P_{3}\), this point produces acceleration for \(0<\lambda <0.82\) and phantom-type characteristics for \(-0.162<\lambda <0\). In this range, the eigenvalues yield negative non-zero real parts, as shown in Fig. [\ref{fig:p4_stab_kes}], indicating the stable behavior of the point. As this point also features a stable accelerating solution, it is physically viable.
			
			\item \textbf{Point \(P_{5}\):} At this point, both field variables vanish, consequently, the field fractional density becomes zero. The fluid energy density dominates and the effective equation of state becomes \(0\), signifying a state similar to a matter-dominated phase. However, upon determining the Jacobian matrix, the corresponding eigenvalues yield negative values, making it stable.
			
		\end{itemize}
		Based on the above critical point analysis, we see that the points \(P_{3,4}\) produce accelerating and phantom solutions based on the range of \(\lambda\). However, these two points do not signify the same state of the universe for a particular value of \(\lambda\), neither do they exist in the same quadrant of the phase space as the coordinates become different. For example, for \(\lambda = 0.2\), the coordinates of \(P_{3}\) become \(P_{3} = (-1.54, 1.84)\), while \(P_{4}  = (1.30, 1.64)\). One of the major differences can be seen in the sign of the \(x\)-coordinate. 
		Due to the existence of an invariant submanifold, the initial conditions in \(x\), if chosen positive, the system evolves and stabilizes at \(P_{4}\). In terms of phase space, one can say that all the trajectories that have originated in the positive \(x\)-axis will remain bound in the first quadrant provided \(y> 0\).
		
		\begin{figure}[t]
			\centering
			\includegraphics[scale=0.5]{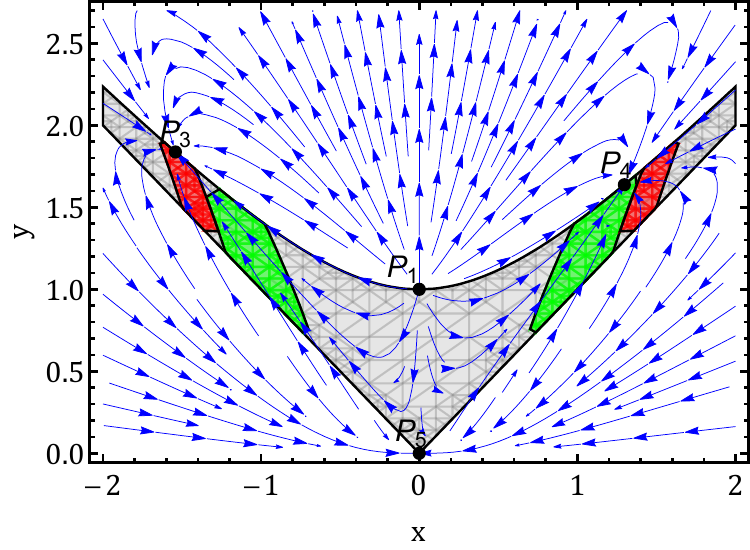}
			\caption{The phase space of $k$-essence field for $\lambda = 0.2$.}
			\label{fig:phase_kes}
		\end{figure}
		\begin{figure}[t]
			\centering
			\includegraphics[scale=0.8]{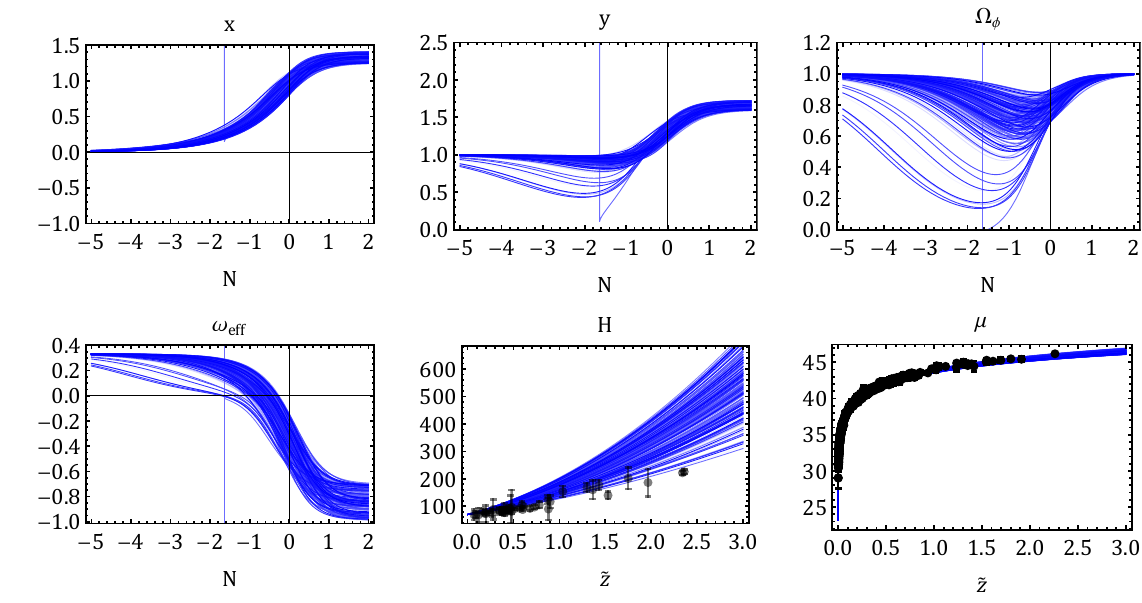}
			\caption{The evolution of cosmological parameters of the k-essence field with power law potential for $H_0= 70 \ \rm Km\  s^{-1} Mpc^{-1}$.}
			\label{fig:evo_kes_entire}
		\end{figure}

		The phase space of the system has been plotted for \(\lambda = 0.2\) in Fig. [\ref{fig:phase_kes}], where the gray region shows the constraint on field energy density, \((0\leq \Omega_{\phi}\leq 1)\). The green region shows the accelerating characteristics, \((-1 \leq \es <-1/3)\), when field energy density dominates \(0.5< \Omega_{\phi} \leq 1\), and the phantom region \((-1.5 <\es < -1)\) is shown in red color. From the phase space, it becomes clear that the trajectories originating from \(P_{1}\) (where near \(P_{1}\) we have radiation-type characteristics) get attracted towards both \(P_{3}\) and \(P_{4}\). Those trajectories whose slopes are positive, \(x>0\) are attracted towards \(P_{4}\), while the trajectories with negative slopes, \(x<0\), get attracted towards \(P_{3}\). Depending on the choice of \(\lambda\), either \(P_{3}\) or \(P_{4}\) shows accelerating characteristics. There are a few trajectories which are not constrained by the Hubble equation and are not attracted towards \(P_{3,4}\); they get attracted to \(P_{5}\), which shows an attractor matter regime. Note that point \(P_{2}\) cannot be considered as a physically viable point, since \(\om\) becomes negative for \(\lambda^2 < 3\). {The trajectory that lies outside the constrained region shown in blue cannot be considered physically viable, as it violates the essential constraints imposed on the system by the Hubble equation, Eq.~\eqref{min_constraint}.} In Fig. [\ref{fig:evo_kes_entire}], we numerically evolved the system by varying the initial conditions and model parameters in the range of
		\begin{equation}
			0.8 \leq x_0 \leq 1.12, \quad  0.1 \leq \Omega_{m0}\leq 0.31, \quad  0.01 \leq \lambda \leq 0.3,
			\label{model1:prameter range}
		\end{equation}
		by generating random numbers. {Here, the subscript zero denotes the initial conditions at \(N=\ti{z}=0\). We have also plotted the evolution of the Hubble parameter $(H)$ in km/s/Mpc and distance modulus, \(\mu\), against redshift, \( a= 1/(1+\ti{z})\), where $\mu$ is defined as
			\begin{equation}
				\mu = 5 \log_{10}(d_{L}) + 25, 
			\end{equation}
			where the luminosity distance is given by:
			\begin{equation}
				d_{L} = c (1+\ti{z}) \int_{0}^{\ti{z}} \frac{1}{H(\ti{z})} d\ti{z},
			\end{equation}
			for the parameter ranges specified above. Here, \(c\) represents the speed of light expressed in km/s. We compare it with the corresponding observational data including 43 OHD and 1701 Pantheon+SH0ES data sets \cite{Cao:2021uda, Pan-STARRS1:2017jku}. 
			Note that, here, we vary \(\Omega_{m0}\) instead of \(y_0\), since one can get \(y_0\) using the Hubble constraint equation as,
			\begin{equation}
				y_0 = \sqrt{1-\Omega_{m0} + x_0^2}\ .
				\label{initial_cond_model_1}
			\end{equation} 
			In the figure, we can see that the initial conditions with \(x_0>0\) produces values of $x$ that remain positive; hence, the system stabilizes at point \(P_{4}\) and generates an accelerating solution. We see that although during the late-time epoch, the field energy density dominates, but in the past epoch, the curves corresponding to the Hubble evolution show deviation from the data points. However, there is no substantial difference that can be accounted for in the distance modulus plot. {In addition to this analysis, it is essential to examine the critical points at infinity to determine if any viable fixed points exist. A comprehensive analysis of these points is provided in Appendix \ref{appen:cric_inf}.} In the subsequent sec. \ref{sec:data_analysis}, we will conduct statistical analysis to obtain the best fit values of the dynamical variables and model parameters. 
		}
		
		\subsection{Stability of the system in the presence of exponential potential}
		
		In this section, we present the dynamical aspects of the \(k\)-essence field in the presence of an exponential type potential. The form of the potential is given by:
		\begin{equation}
			V(\phi) = V_0 e^{\beta H_0 \phi}\, .
		\end{equation}
		Here, $V_0$ and \(H_0\) are dimensional constants with mass dimensions $[M]^{4}$ and $[M]$, respectively, whereas $\beta$ is a dimensionless constant. To study the dynamics corresponding to the present case, we define a new set of dimensionless dynamical variables. These dynamical variables are different from the former case, although in some cases the notations are the same. The variables are:
		\begin{equation}
			x = \dot{\phi}, \quad y = \phi H_0, \quad z = \frac{H - H_0}{H+ H_0}\,,
			\label{dyn_variab_exponential}
		\end{equation}
		These are the primary variables required to close the system. Here, \(H_0\) denotes the present value of the Hubble parameter. In the previous case, where the potential was considered of the inverse square type, the derivative of the potential, denoted as \(\lambda\), turned out to be constant. As a result, only two variables were required to close the system. In contrast, the exponential type potential does not yield a constant \(\lambda\) and thus requires more variables to close the system. Therefore, we have chosen the variables such that the dynamics of the system can be analyzed with a minimal number of dimensionless dynamical variables. It should be noted that the variables \((x,y)\) can take any value, as the range is given by:
		\begin{equation}
			-\infty <x< \infty, \quad -\infty <y< \infty,
			\label{dyn_var_range}
		\end{equation}
		whereas we have compacted the variable \(z\) in the range given by:
		\begin{equation}
			z=
			\begin{cases}
				-1 & H \ll H_0,\\
				0 & H =H_0,\\
				1 & H \to + \infty\ .
			\end{cases}
		\end{equation}
		The secondary variables, for instance, the fractional densities of the field and fluid, become:
		\begin{equation}
			\Omega_{\phi} := \frac{\kappa^2 \rho_{\phi}}{3 H^2} = \frac{\mathcal{M}_0}{3} \left(\frac{1-z}{1+z}\right)^{2} e^{\beta y} \left(\frac{3x^4}{4} - \frac{x^2}{2}\right), \quad \Omega_{m} = \frac{\kappa^2 \rho_m}{3 H^2}\,.
			\label{density_expo}
		\end{equation}
		Here, \(\mathcal{M}_0 \equiv \frac{\kappa^2 V_0}{H_0^2}\) is a dimensional constant. The Hubble equation puts a constraint on \(\Omega_{m}\), which must be in the range given by:
		\begin{equation}
			\Omega_{m} = 1 - \Omega_{\phi}, \quad 0 \le \Omega_{m} \le 1, \quad 0 \le \Omega_{\phi} \le 1.
			\label{constraint_eq_hub_expo}
		\end{equation}
		Because of the Friedmann constraint relation, the variables \((x,y)\) cannot take any values. Solutions that do not satisfy the above relation cannot be considered physically viable. Similarly, the effective equation of state (EoS) can be expressed in terms of the dynamical variable as: 
		\begin{equation}
			\omega_{\rm eff} = \frac{\mathcal{M}_0}{3} \left(\frac{1-z}{1+z}\right)^{2} e^{\beta y} \left(\frac{x^4}{4}-\frac{x^2}{2}\right).
		\end{equation}
		The autonomous system of equations can be constructed in the predefined variables as:
		\begin{eqnarray}
			x' & =& \frac{-3 x^3+\frac{\beta  \left(\frac{x^2}{2}-\frac{3 x^4}{4}\right) (1-z)}{z+1}+3 x}{3 x^2-1}, \\
			y' & =& \frac{x (1-z)}{z+1}, \label{y_prime_expo}\\
			z' & =& -\frac{3 ((1-z) (z+1)) \left(\frac{1}{3} \mathcal{M}_0 \left(\frac{x^4}{4}-\frac{x^2}{2}\right) \left(\frac{1-z}{z+1}\right)^2 \exp (\beta  y)+1\right)}{4}\, .
		\end{eqnarray}
		Here, \(()' \equiv \frac{d()}{H dt}\), where \(H dt = dN\). Since the minimum number of dynamical variables is three, the corresponding phase space becomes 3D. Although the system of equations is consistent at \(z = 1\), i.e., \(H \gg H_0\), one may notice the pathology of the dynamical system at \(z = -1\), i.e., \(H \ll H_0\). To study the dynamics corresponding to this coordinate, we use the following standard trick usually used in dynamical systems analysis by redefining the time variable as \cite{Bahamonde:2017ize}:
		\begin{equation}
			dN \to (1+z) d\tilde{N}.
			\label{redef_n_variable}
		\end{equation}
		With this redefinition, the autonomous system of equations becomes non-divergent. The redefinition does not alter the dynamics of the system as long as the system is far from this critical point. However, in the data analysis section, to constrain the parameters, we shall evolve the system against the original time variable \(dN = d \log(1/(1+\tilde{z}))\), where \(\tilde{z}\) is the redshift. The experimental data available to us is in the redshift range \(\tilde{z} \in [0.001, 2.5]\), whereas \(( \tilde{z} \to -1)\) shows the future state of the universe.
		
		\begin{table}[h!]
			\centering
			\begin{tabular}{ccccc}
				\hline
				Points & $(x,y,z)$ & $\Omega_{\phi}$ & $\Omega_{m}$ & $\omega_{\rm eff}$ \\
				\hline
				$P_{1,2}$ & $(0, \text{Any}, \mp 1)$ & 0 & 1 & 0 \\

				\hline
				$P_{3,4}$ & $(\mp 1, \text{Any}, 1)$ & 0 & 1 & 0\\
				
				\hline
				\hline
			\end{tabular}
			\caption{The nature of critical points corresponding to an exponential potential.}
			\label{tab:cric_expo}
		\end{table}
		{To analyze the system’s dynamics, we first determine the system's critical points using $\tilde{N}$ and examine their stability. The critical points are presented in Table [\ref{tab:cric_expo}]. In the table - Any - shows that the $y$ coordinate can be chosen arbitrarily, the critical point analysis does not specify any particular values for the $y$-coordinate of the critical points. The system reveals four critical points where the matter fractional energy density dominates while the field fractional energy density vanishes. This results in an effective equation of state (EoS) that vanishes, indicating a matter-dominated regime.
			
			Given the complexity of the system, it is unable to generate critical points that can directly constrain the model parameters \( \beta \) and \( \mathcal{M}_0 \), as was possible in previous cases. Consequently, a purely numerical approach is required to evolve the system with respect to \( \tilde{N} \) by varying initial conditions and model parameters to identify the best range of initial conditions that can replicate the different epochs of the universe.
			
			Before moving on to the numerical evolution, we address a key concern for readers. The ranges for our dynamical variables are specified in Eq.~\eqref{dyn_var_range}. This range implies the potential existence of critical points at infinity, which could shed light on accelerating critical points. To address this, we utilize the Poincaré transformation for one of the variables, say \( \bar{y} = y/\sqrt{1 + y^2} \). Since \( y \) can range from \( -\infty \) to \( +\infty \), this transformation maps these infinities to the interval \( -1 \) to \( +1 \), thereby compactifying the dynamical variable. With both \( z \) and \( \bar{y} \) confined within finite ranges, there is no need to compactify the variable \( x \), as the Hubble constraint enforces valid physical solutions. Following this compactification approach, we can identify the critical points at infinity. In Appendix \ref{appen:2}, we demonstrate that the transformed variables do not exhibit any additional critical points beyond the current case, and, upon numerical simulation, the system dynamics remain consistent with our current scenario.
			
			It is also worth noting that this compactification exercise can be avoided by a different line of reasoning. Although the dynamical variables can theoretically take any values, the Hubble constraint requires that the fractional energy densities (field or fluid) remain within \( 0 \leq \Omega_{\phi} \leq 1 \). From Eq.~\eqref{density_expo}, we observe that the field energy density parameter depends on all variables, with \( z \) already compactified. Therefore, the remaining values for \( (x, y) \) must only yield finite densities. Any critical point that fails to satisfy this constraint cannot be considered physically viable.}

		In the present case a conventional critical point analysis does not shed much light on the accelerating expansion phase of the universe and consequently we proceed with direct numerical evolution of the system using the autonomous equations. We take a different route to study the dynamics of the system. We skip the critical point analysis and numerically evolve the system by varying the parameters and initial conditions to obtain the dynamics of the system. We vary the appropriate variables within this range by generating random numbers as given by:  
		\begin{equation}
			1.01 < x_0 < 1.25, \quad 0.3 < \Omega_{m0} < 0.31, \quad z_0 = 0, \quad 5 < \mathcal{M}_0 < 9, \quad -0.2 < \beta < 0.2.
			\label{model2:prameter range}
		\end{equation}
		Here, the subscript \(0\) indicates the value at the present epoch, for instance \(x_0 \equiv x(\ti{N}=0)\). We have fixed \(z_0 = 0\) due to its definition in Eq.~\eqref{dyn_variab_exponential}. In this approach, we choose to vary \(\Omega_{m0}\) instead of \(y_0\) since, by following the Hubble constraint equation \(\Omega_{m0} = 1 - \Omega_{\phi0}\), one obtains the initial conditions for \(y_0\). Note that this technique has also been adopted in the data analysis section \ref{sec:data_analysis}.
		
		\begin{figure}[t]
			\centering
			\includegraphics[scale=0.8]{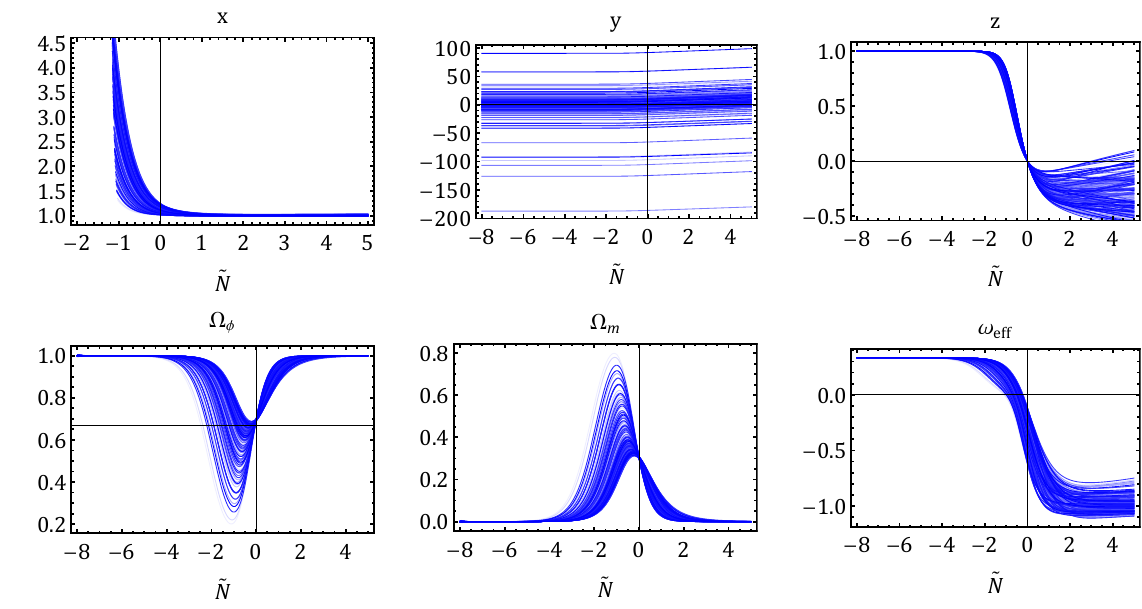}
			\caption{The evolution of cosmological parameters of the k-essence field for an exponential potential.}
			\label{fig:evo_kes_entire_expo}
		\end{figure}
		
		\begin{figure}[t]
			\centering
			\includegraphics[scale=0.8]{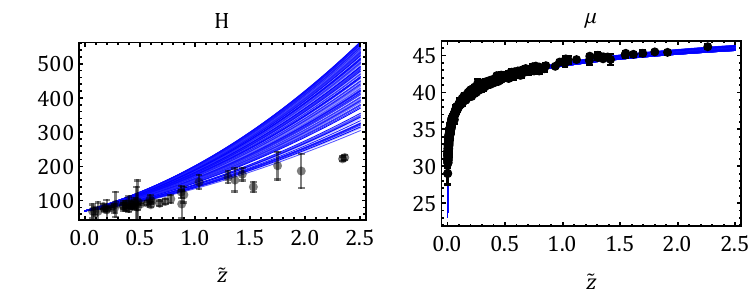}
			\caption{The evolution of cosmological observable \(H\) in km/s/Mpc and distance modulus \(\mu\) against redshift $\ti{z}$ for the exponential potential. Here we considered the value of \(H_0 = 70\) km/s/Mpc. }
			\label{fig:kes_expo_hmu}
		\end{figure}

		The summary of the evolution of cosmological parameters is illustrated in Fig. [\ref{fig:evo_kes_entire_expo}]. From the evolution of fractional densities of the field and fluid, it can be concluded that for a certain number of e-folds, \(-3 < \ti{N} < -0.5\), the fluid energy density dominates over the field energy density, and the corresponding effective EoS \((\omega_{\rm eff})\) vanishes. This indicates that the matter phase becomes an attractor phase for the \(k\)-essence field with an exponential potential over a wide parameter range. In the far past epoch, we see that \(\omega_{\rm eff}\) saturates to \(1/3\) and the corresponding field energy density dominates. This suggests that the \(k\)-essence field with an exponential potential can produce a similar effect as a radiation fluid. As the system evolves towards the current epoch \(\ti{N} = 0\), the field fractional density starts dominating, and the corresponding effective EoS decreases towards negative values, exhibiting a tracking behavior. The equation of state of the field becomes smaller than the EoS of the background fluid, causing the system to accelerate, which exhibits dark energy behavior. The dynamical variable \(x\) corresponding to this phase nearly saturates to \(1\). Throughout the evolution, the other dynamical variables do not diverge but take different values as the model parameters change.
		
		As the system enters the far future \(\ti{N }> 0\), the effective EoS \(\omega_{\rm eff}\) for some combination of parameters even crosses \(-1\), exhibiting phantom behavior, and the corresponding dynamical variable \(z\) approaches \(-1\). However, an additional interesting behavior can be observed when some curves of \(\omega_{\rm eff}\) move towards positive values, hinting at a change of phase from an accelerating to a decelerating epoch. This scenario is also confirmed from the best fit obtained in Fig. [\ref{fig:evo_kes_exp_best}] for various combination of data sets. At this stage, the field density remains dominant over the fluid density.
		
		Hence, from this analysis, it can be concluded that the \(k\)-essence field with an exponential potential can generate different phases of the universe, namely:
		\begin{equation*}
			\text{Radiation} \longrightarrow \text{Matter} \longrightarrow \text{Acceleration} \overset{\text{or}}{\longrightarrow }\text{Phantom} \longrightarrow \text{Deceleration}.
		\end{equation*}
		Additionally, we have plotted the Hubble parameter \(H\) and distance modulus \(\mu\) in Fig. [\ref{fig:kes_expo_hmu}], with the redshift \(\tilde{z}\) assuming \(H_0 = 70\) km/s/Mpc. It can be seen that the current model produces curves close to the observational data points. However, the best fit values of the parameters can only be obtained by accounting for MCMC with different data sets, which will be presented in the next section.

		\section{Data Analysis}
		\label{sec:data_analysis}
		
		Our subsequent objective involves performing data analysis to constrain the parameter space of the model. In order to achieve this goal, we need to derive an analytical expression for the Hubble parameter as a function of redshift. While this process is straightforward for the $\Lambda$CDM model, it poses challenges for $k$-essence models due to the inability to obtain an analytical solution to the Friedmann equation. Therefore, we opted for a dynamical systems approach, solving the coupled differential equations outlined in Eq. (\ref{eq:2}) to capture the evolution of the Hubble parameter. 
		{ Here, we  constraint the model parameters, utilizing the following data sets: 
			\begin{itemize}
				
				\item \textbf{The Hubble datasets:} This dataset contains $31$ points of Hubble parameter which are derived from the differential age method \cite{Yu:2017iju}. We refer to this dataset as CC. The chisquare (\(\chi^2\)) for this dataset is computed using the expression,
				\begin{equation}
					\chi^2 = \sum_{i=1}^{31} \left(\frac{H_{i \ \rm obs} - H_{i \  \rm Model}}{\sigma_{i}}\right)^{2},
				\end{equation}
				{where $H_{i \ \rm obs}$ is the Hubble parameter observed at redshift $z_i$, $\sigma_i$ is the error in the measurement and $H_{i \ \rm Model}$ is the corresponding model prediction.}
				
				\item \textbf{Pantheon+SH0ES data:} This data set contains most recent compilation of Type Ia supernovae (SNe Ia) from the Pantheon+SH0ES dataset \cite{brout2022pantheon+, scolnic2022pantheon+}. The pantheon+ compilation include 1701 SNe Ia light curves of 1550 supernovae observed in the redshift span $0\leq \tilde{z} \leq 2.3$ \cite{Pan-STARRS1:2017jku}. It is pertinent to remind the reader that in our notation the cosmological redshift is specified by $\tilde{z}$ and not simply $z$ as is the conventional practice.
				We refer to this data set as PP throughout the text. The observable in case of SNe Ia is the distance modulus ($\mu$) which can be computed theoretically using the expression,
				\begin{equation}
					\label{eq:m}
					\mu (\tilde{z}) = 5\,\log_{10}\left[\frac{d_L(\tilde{z})}{Mpc}\right] + 25,
				\end{equation}
				The luminosity distance $d_L$ in a flat FLRW universe can be calculated as \cite{Brout:2022vxf},
				\begin{equation}
					\label{eq:lum}
					d_L(\tilde{z}) = c(1+\tilde{z}) \int_0^{\tilde{z}} \frac{dz'}{H(z')},
				\end{equation}
				where $c$ is the speed of light in vacuum expressed in the Km/s unit. For the SNe Ia, we compute the $\chi^2$ using the expression,
				\begin{equation}
					\label{eq:chi2sne}
					\chi^2_{\text{SNe Ia}} = \Delta \vec{D}^T C^{-1}_{\rm stat+syst} \Delta \vec{D},
				\end{equation}
				where $C_{\rm stat+syst} = C_{\rm stat} + C_{\rm syst}$ is the combined covariance matrix of systematic and statistical uncertainties and $\vec{D}$ is the vector of SNe Ia distance modulus calculated as 
				\begin{equation}
					\label{eq:DM}
					\Delta \vec{D} = \mu(\tilde{z}_i) - \mu_{th}(\tilde{z}_i,H_0,\Omega_{m_0},y_0).
				\end{equation}
				
				\item \textbf{DES data:} {The DES data set include 1829 SNe Ia in the redshift range $0.01 < \tilde{z} < 1.13$ \cite{abbott2024dark}.}
				
				\item \textbf{BAO data:} {The data set contains 8 data points from Sloan Digital Sky Survey (SDSS) \cite{eBOSS:2020yzd, dawson2016sdss} called as BAO in the following discussion and $7$ data points from Dark Energy Spectroscopic Instrument (DESI) \cite{DESI:2024mwx, levi2019dark, moon2023first} called as DESBAO in the following discussion, which are tabulated in Tab. [\ref{tab:bao_measurements}].}  The $\chi^2$ is defined as:
				\begin{equation}
					\chi^2_{\rm BAO} = \sum_{i=1}^{N} \left(\frac{X_i}{\sigma_{i}}\right)^2
				\end{equation}
				where $X$ is given by: \begin{equation}
					X_i = \text{Data}_i - \text{Model}_i\ ,
				\end{equation}
				Here, the observables are for instance $D_{M}(z)$, which is the comoving angular-diameter distance defined as $D_{M}(\tilde{z}) =  \int_{0}^{\tilde{z}} c \frac{dz^\prime}{H(z^\prime)} $, the Hubble distance is $D_H(\tilde{z}) = c/H(\tilde{z})$ and the spherically-averaged distance $D_{V}(\tilde{z})$ is called dilation scale defined as:
				\begin{equation}
					D_{V}(\tilde{z}) \equiv \left({\tilde{z} \ D_M^2(\tilde{z})  D_H}\right)^{1/3}\ .
				\end{equation} 
				The parameter \(r_d\) shows the distance traveled by sound waves between the end of inflation and the decoupling of baryons from photons after recombination,
				\begin{equation}
					r_d  = \int_{\tilde{z}_d = 1020}^{\infty} \frac{c_s(z^\prime)}{H(z^\prime)} dz^\prime,
				\end{equation}
				where \(c_s\) denotes the sound speed and \(\tilde{z}_d\) is the redshift of the drag epoch. \(c_s\) depends on the baryon and photon density which can be calculated as: 
				\begin{equation}
					c_s(\tilde{z}) = \frac{c}{\sqrt{3(1 + \frac{3 \rho_{B}}{4 \rho_{\gamma}})}}\, .
				\end{equation}
				However, in this study, we treat $r_d$ as a free parameter which we have constrained using the present combination of data sets. Note that to calculate the $\chi^2_{\rm BAO}$, we used \(\chi^2_{\rm tot} = \frac12 \left(\chi^{2}_{D_M/r_d}  +\chi^{2}_{D_H/r_d}+ \chi^{2}_{D_V/r_d}\right)\) \cite{eBOSS:2020yzd}.

				\begin{table}[h!]
					\tiny
					\centering
					
					\begin{tabular}{|c|c|c|c|c|c|c|c|c|}
						\hline
						\multicolumn{9}{|c|}{BAO measurement} \\
						\hline
						
						$\tilde{z}_{\rm eff}$ & 0.15 & 0.38 & 0.51 & 0.70& 0.85 & 1.48 & 2.33& 2.33 \\
						
						\hline
						$D_V(\tilde{z})/r_d$ & $4.47 \pm 0.17$ & & & &$18.33^{+0.57}_{-0.62}$  & &&\\
						\hline
						
						$D_M(\tilde{z})/r_d$ & & $10.23 \pm 0.17$ & $13.36 \pm 0.21$ & $17.86 \pm 0.33$ & & $30.69 \pm 0.80$ & $37.6 \pm 1.9$ & $37.3 \pm 1.7$ \\
						\hline
						
						$D_H(\tilde{z})/r_d$ & & $25.00 \pm 0.76$ & $22.33 \pm 0.58$ & $19.33 \pm 0.53$ & & $13.26 \pm 0.55$ & $8.93 \pm 0.28$ & $9.08 \pm 0.34 $\\
						\hline
						\hline
						\multicolumn{9}{|c|}{DESBAO measurements}\\
						\hline
						$\tilde{z}_{\rm eff}$ & 0.295 & 0.510 & 0.706 & 0.930 & 1.317 & 1.491 & 2.330 & \\
						\hline
						$D_V(\tilde{z})/r_d$ & $7.93 \pm 0.15$ &  & && &$26.07 \pm 0.67$ && \\
						\hline
						
						$D_M(\tilde{z})/r_d$ & & $13.62 \pm 0.25$ & $16.85 \pm 0.32$ & $21.71 \pm 0.28$ & $27.79 \pm 0.69$ & & $39.71 \pm 0.94$ &  \\
						\hline
						
						$D_H(\tilde{z})/r_d$ & & $20.98 \pm 0.61$ & $20.08 \pm 0.60$ & $17.88 \pm 0.35$ & $13.82 \pm 0.42$ &  & $8.52 \pm 0.17$ & \\
						\hline
						\hline
						
					\end{tabular}
					\caption{BAO data.}
					\label{tab:bao_measurements}
				\end{table}
				
			\end{itemize}
		}
	We use the Bayesian parameter inference which relies on the Bayes theorem, offering a gratifying description to compute the posterior probability of a particular parameter of interest given the data (D) and model (M). Mathematically, it can be expressed as \cite{Trotta:2008qt, N:2021qlo},
	\begin{align}
		\label{eq:Bayes}
		P(\theta|D,M) = \frac{P(D|\theta,M)P(\theta|M)}{P(D|M)},
	\end{align}
	where $P(\theta|D,M)$ denotes the posterior distribution of the model parameters, $P(D|\theta,M)$ stands for the  likelihood, $P(\theta|M)$ is the prior probability distribution. Meanwhile, $P(D|M)$ serves as a normalization factor representing the evidence of the model. Although $P(D|M)$ is insignificant from the parameter estimation, it plays the central role in the model selection. The prior probability incorporate any existing knowledge about the model parameters before the data acquisition. The selection of a prior relies on the expertise of the researcher in the field and quality of judgement. Nevertheless, once the prior is set, iterative application of the Bayes theorem converges to a common posterior \cite{Padilla:2019mgi}. The likelihood is assumed to be Gaussian,
	\begin{equation}
		\label{eq:gauss}
		P(D|\theta,M) \equiv \exp\left(\frac{-\chi^2(\theta)}{2}\right).
	\end{equation}

	The best fit values of the model parameters are those that minimize the total $\chi^2$.
	In order to obtain the 1D and 2D posterior probability distribution of the model parameters, we marginalize over the parameters except the parameters under consideration. {Information criteria, such as the Akaike Information Criterion (AIC) \cite{1100705} and the Bayesian Information Criterion (BIC) \cite{c4048c8f-6ca9-3965-96a3-653ab8996955}, are utilized to compare how well $k$-essence models fit the data in comparison to the $\Lambda$CDM model.}
	The AIC and BIC are defined as \cite{Trotta:2008qt},
	\begin{align}
		\rm{AIC} = -2 \ln \mathcal{L}_{\rm max} + 2 k,\\
		\rm{BIC} = -2 \ln \mathcal{L}_{\rm max} 
		+ k \ln N
	\end{align}
	where $k$ is the number of independent parameters present in the model and $N$ represent the total number of data points and \(\mathcal{L}_{\rm max} \) represents the maximum likelihood. Performing parameter inference involve careful selection of priors for the model parameters. 
	\begin{table}[h!]
		\centering
		\small
		\begin{tabular}{lcccrr}
			\hline
			\multicolumn{6}{c}{\texttt{Flat $\Lambda$CDM model}} \\
			\hline
			Dataset & $H_0$ & $\Omega_m$ & $r_d$ & AIC & BIC\\
			\hline
			Priors & [30,100] & [0,0.6] & [100, 300]& & \\
			\hline
			CC+BAO & $69.8\pm 1.9$ & $0.297^{+0.017}_{-0.019}$ & $144.1\pm 3.6$ & 29.68 & 34.74 \\
			CC+BAO+DESBAO & $69.5\pm 1.8$ & $0.302^{+0.011}_{-0.013}$ & $145.1^{+3.4}_{-3.8}$ & 45.11 & 50.66 \\
			CC+BAO+DESBAO+DES & $69.93\pm 0.24$ & $0.3185\pm 0.0098$ & $142.63\pm 0.85$ & 1691.39 & 1708.00 \\
			CC+BAO+DESBAO+PP+DES & $72.56\pm 0.14$ & $0.2691\pm 0.0068$ & $142.29\pm 0.85$ & 3749.39 & 3767.94 \\
			CC+BAO+DESBAO+PP & $73.32\pm 0.17$ & $0.315\pm 0.010$ & $136.38\pm 0.98$ & 1811.50 & 1827.90 \\
			CC+BAO+DES & $69.78\pm 0.28$ & $0.325\pm 0.012$ & $141.7\pm 1.1$ & 1675.42 & 1692.02 \\
			CC+BAO+DES+PP & $72.73\pm 0.15$ & $0.2582\pm 0.0077$ & $141.8\pm 1.1$ & 3726.06 & 3744.60 \\
			CC+BAO+PP & $73.26\pm 0.18$ & $0.320\pm 0.012$ & $135.5\pm 1.2$ & 1795.95 & 1812.34 \\
			CC+DESBAO & $69.5\pm 1.8$ & $0.304\pm 0.016$ & $145.7\pm 3.6$ & 31.86 & 36.85 \\
			CC+DESBAO+DES & $69.79\pm 0.26$ & $0.325\pm 0.012$ & $142.7\pm 1.1$ & 1676.92 & 1693.52 \\
			CC+DESBAO+DES+PP & $72.66\pm 0.15$ & $0.2630\pm 0.0074$ & $143.6\pm 1.1$ & 3732.29 & 3750.83 \\
			CC+DESBAO+PP & $73.25\pm 0.18$ & $0.320^{+0.011}_{-0.012}$ & $136.4\pm 1.3$ & 1797.49 & 1813.88 \\
			
			\hline
			\hline
		\end{tabular}
		
		\caption{Summary of best-fit values for the parameters at the 68\% confidence level for  $\Lambda$CDM. }
		
		\label{tab:lcdm}
	\end{table}

	\begin{table}[h!]
		\centering
		
		\scriptsize
		\resizebox{\textwidth}{!}{%
			\begin{tabularx}{\textwidth}{lXXXXXrr}
				\hline
				\multicolumn{8}{c}{\texttt{Model I: Inverse Square Potential}}\\
				\hline
				Dataset & $H_0$ & $\Omega_m$ & $r_d$ & $x_0$ & $\lambda$ & AIC & BIC \\
				\hline
				Priors & [30,100] & [0,0.8] & [100, 300] & [0.6, 2.4] & [-1.5, 2.0] & & \\
				\hline
				CC+BAO & $66.9\pm 2.4$ & $0.196^{+0.11}_{-0.057}$ & $144.7\pm 3.6$ & $1.097^{+0.054}_{-0.060}$ & $0.34^{+0.19}_{-0.16}$ & 30.92 & 39.36 \\
				CC+BAO+DESBAO & $67.5\pm 2.1$ & $0.185^{+0.12}_{-0.067}$ & $145.2\pm 3.7$ & $1.125\pm 0.044$ & $0.26^{+0.15}_{-0.12}$ & 46.28 & 55.53 \\
				CC+BAO+DESBAO+DES & $68.99\pm 0.36$ & $0.180^{+0.12}_{-0.066}$ & $142.76\pm 0.86$ & $1.136\pm 0.021$ & $0.234^{+0.071}_{-0.062}$ & 1685.91 & 1713.59 \\
				CC+BAO+DESBAO+PP+DES & $72.84\pm 0.21$ & $0.191^{+0.084}_{-0.046}$ & $142.07\pm 0.83$ & $1.243\pm 0.013$ & $-0.051^{+0.054}_{-0.061}$ & 3746.29 & 3777.20 \\
				CC+BAO+DESBAO+PP & $72.64\pm 0.24$ & $0.157\pm 0.090$ & $135.72\pm 0.98$ & $1.139^{+0.023}_{-0.019}$ & $0.251^{+0.066}_{-0.058}$ & 1802.39 & 1829.73 \\
				CC+BAO+DES & $68.95\pm 0.36$ & $0.195^{+0.11}_{-0.052}$ & $142.3\pm 1.1$ & $1.128\pm 0.021$ & $0.255^{+0.072}_{-0.063}$ & 1670.43 & 1698.09 \\
				CC+BAO+DES+PP & $72.84\pm 0.21$ & $0.176^{+0.091}_{-0.049}$ & $141.4\pm 1.1$ & $1.242\pm 0.014$ & $-0.018\pm 0.06$ & 3727.72 & 3758.63 \\
				
				CC+BAO+PP & $72.61\pm 0.25$ & $0.175^{+0.12}_{-0.059}$ & $135.3\pm 1.2$ & $1.130\pm 0.023$ & $0.275^{+0.071}_{-0.058}$ & 1786.94 & 1814.25 \\
				
				CC+DESBAO & $68.8^{+2.4}_{-2.7}$ & $0.175^{+0.12}_{-0.073}$ & $145.6^{+3.4}_{-3.9}$ & $1.173^{+0.052}_{-0.066}$ & $0.13^{+0.21}_{-0.15}$ & 34.36 & 42.68 \\
				CC+DESBAO+DES & $69.09\pm 0.36$ & $0.149^{+0.065}_{-0.13}$ & $143.2\pm 1.2$ & $1.148^{+0.023}_{-0.020}$ & $0.221^{+0.076}_{-0.058}$ & 1673.30 & 1700.96 \\
				CC+DESBAO+DES+PP & $72.86\pm 0.21$ & $0.119^{+0.056}_{-0.094}$ & $143.1\pm 1.1$ & $1.254^{+0.014}_{-0.012}$ & $-0.016^{+0.069}_{-0.052}$ & 3727.20 & 3758.10 \\
				CC+DESBAO+PP & $72.70\pm 0.24$ & $0.127^{+0.051}_{-0.12}$ & $136.4\pm 1.2$ & $1.150^{+0.024}_{-0.020}$ & $0.241^{+0.066}_{-0.054}$ & 1789.57 & 1816.88 \\
				\hline
				\hline
			\end{tabularx}
		}
		\caption{Summary of best-fit values for the parameters at the 68\% confidence level for Model I.}
		\label{tab:model1_confidence}
	\end{table}
	
	\begin{table}[h!]
		\centering
		\scriptsize
		\resizebox{\textwidth}{!}{%
			\begin{tabularx}{\textwidth}{lXXXXXrr}
				
				\hline
				\multicolumn{8}{c}{\texttt{Model II: Exponential potential with fixed parameter: $\mathcal{M}_0 = 8$.}}\\
				\hline
				Dataset & $H_0$ & $\Omega_{m}$ & $r_d$ & $x_0$ & $\beta$ & AIC & BIC \\
				\hline
				Priors & $[30,100]$ & [0,0.6] & [100, 300] & [0.7, 2.2] & [-2.5, 2.5] &  & \\
				\hline
				CC+BAO & $66.3\pm 2.5$ & $0.187^{+0.12}_{-0.062}$ & $144.8\pm 3.7$ & $1.071^{+0.032}_{-0.038}$ & $-1.25^{+0.52}_{-0.61}$ & 31.08 & 39.52 \\
				CC+BAO+DESBAO & $67.0\pm 2.1$ & $0.166\pm 0.092$ & $145.3\pm 3.6$ & $1.061\pm 0.029$ & $-0.97^{+0.41}_{-0.47}$ & 46.21 & 55.46 \\
				CC+BAO+DESBAO+DES & $68.93\pm 0.37$ & $0.174^{+0.12}_{-0.075}$ & $142.79\pm 0.87$ & $1.047^{+0.019}_{-0.022}$ & $-0.73\pm 0.21$ & 1685.90 & 1713.58 \\
				CC+BAO+DESBAO+PP+DES & $72.84\pm 0.21$ & $0.184^{+0.091}_{-0.051}$ & $142.07\pm 0.87$ & $1.009^{+0.011}_{-0.017}$ & $0.14^{+0.19}_{-0.17}$ & 3747.10 & 3778.01 \\
				CC+BAO+DESBAO+PP & $72.60\pm 0.25$ & $0.151^{+0.079}_{-0.11}$ & $135.73\pm 0.97$ & $1.052^{+0.020}_{-0.018}$ & $-0.78\pm 0.19$ & 1802.61 & 1829.95 \\
				CC+BAO+DES & $68.92\pm 0.38$ & $0.191^{+0.11}_{-0.055}$ & $142.3\pm 1.1$ & $1.045^{+0.015}_{-0.021}$ & $-0.77\pm 0.22$ & 1670.49 & 1698.15 \\
				CC+BAO+DES+PP & $72.84\pm 0.21$ & $0.167^{+0.098}_{-0.055}$ & $141.5\pm 1.1$ & $1.013^{+0.012}_{-0.017}$ & $0.04\pm 0.18$ & 3728.46 & 3759.36 \\
				CC+BAO+PP & $72.57\pm 0.25$ & $0.163^{+0.11}_{-0.088}$ & $135.3\pm 1.2$ & $1.052^{+0.018}_{-0.020}$ & $-0.87^{+0.19}_{-0.22}$ & 1787.15 & 1814.46 \\
				CC+DESBAO & $67.8\pm 2.6$ & $0.153\pm 0.093$ & $145.7^{+3.4}_{-3.8}$ & $1.051\pm 0.034$ & $-0.69^{+0.53}_{-0.63}$ & 34.19 & 42.50 \\
				CC+DESBAO+DES & $69.02\pm 0.37$ & $0.141^{+0.064}_{-0.12}$ & $143.3\pm 1.1$ & $1.051^{+0.022}_{-0.016}$ & $-0.70^{+0.19}_{-0.23}$ & 1673.09 & 1700.75 \\
				CC+DESBAO+DES+PP & $72.87\pm 0.21$ & $0.113^{+0.047}_{-0.098}$ & $143.1\pm 1.1$ & $1.020^{+0.018}_{-0.010}$ & $0.04^{+0.16}_{-0.21}$ & 3727.83 & 3758.73 \\
				CC+DESBAO+PP & $72.65\pm 0.25$ & $0.116^{+0.044}_{-0.11}$ & $136.4\pm 1.2$ & $1.056^{+0.022}_{-0.012}$ & $-0.77^{+0.17}_{-0.21}$ & 1789.38 & 1816.69 \\
				
				\hline
				\hline
			\end{tabularx}
		}
		\caption{Summary of best-fit values for the parameters at the 68\% confidence level for Model II.}
		\label{tab:mode2_confidence}
	\end{table}

	
	\begin{figure}[h!]
			\centering
			\subfloat[\label{fig:comp_evo_kes_power}]{	\includegraphics[scale=0.75]{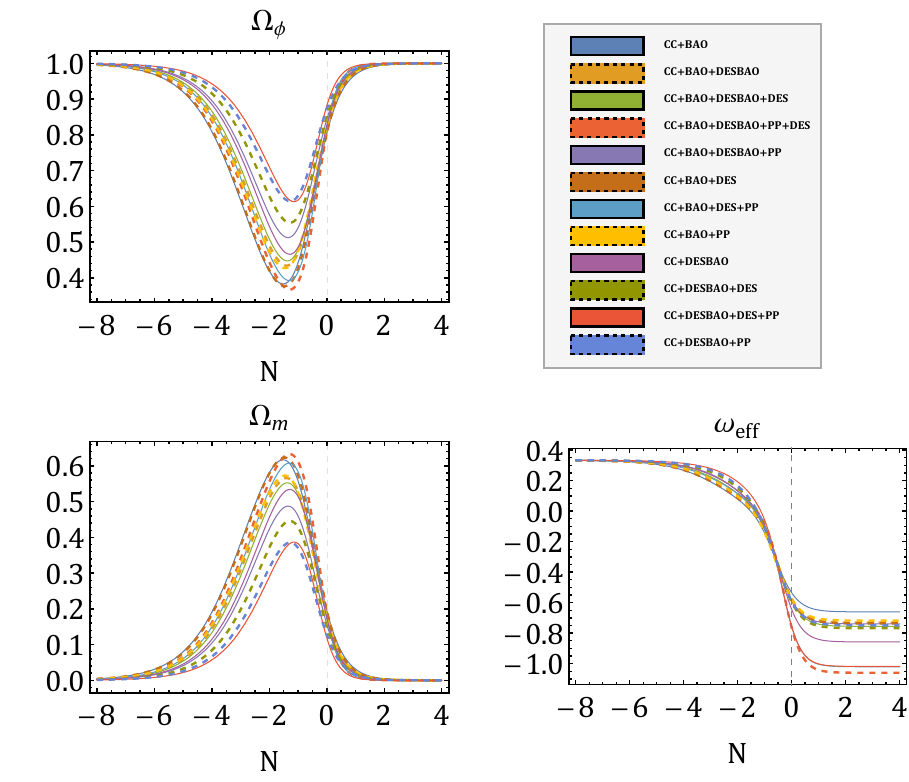}}
			
			\subfloat[\label{fig:evo_kes_exp_best}]{\includegraphics[scale=0.75]{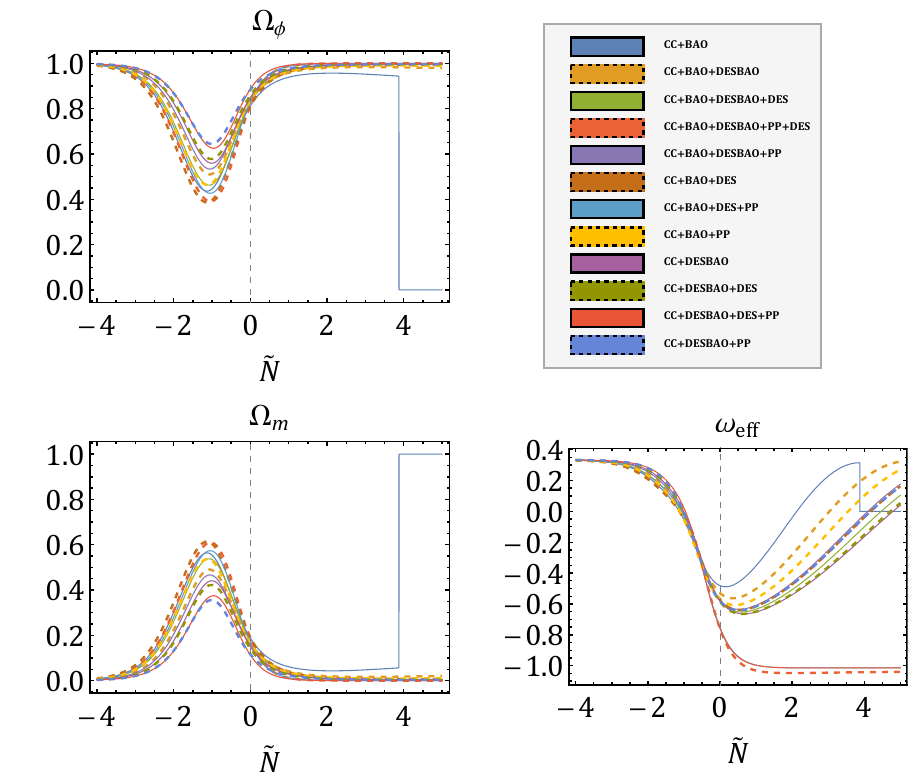}}
			
			\caption{Evolution of the $k$-essence field for (a) Inverse square potential and (b) exponential potential corresponding to the best fit values of model parameters obtained from the data analysis.   }
		\end{figure}

		We utilized a well tested and openly accessible Python implementation of the affine-invariant ensemble sampler for Markov Chain Monte Carlo (MCMC), as proposed by Goodman and Weare, known as emcee for the Bayesian analysis. More comprehensive information about emcee and its implementation can be found in ref. \cite{Foreman_Mackey_2013}. 
		
		{Marginalized 1D and 2D posterior distribution of the dynamical variables and model parameters for the $\Lambda$CDM model, Model I and Model II for various data sets are shown in Fig. [\ref{fig:triangle_plot}]. The mean and standard deviation of the distribution are considered as the best-fit value and error corresponding to each model parameters, which are summarized in Tabs. [\ref{tab:lcdm}, \ref{tab:model1_confidence}, \ref{tab:mode2_confidence}] respectively.}
		{In the table, we provided the dynamical variables and model parameters corresponding to Model I and Model II and specified the corresponding prior ranges. These prior ranges were selected based on the analysis conducted in the previous dynamical stability section. Note that the parameter ranges specified in Eqs. \eqref{model1:prameter range} and \eqref{model2:prameter range} are extended in Tabs. [\ref{tab:model1_confidence}, \ref{tab:mode2_confidence}].}
		
		{For Model I, which refers to the \(k\)-essence field with inverse square potential, there are mainly two primary variables, \((x,y)\), whose initial conditions \((x_0, y_0)\) must be varied. However, in the table, we have listed \(x_0\) and \(\Omega_{m0}\) instead of \(y_0\). The reason for this is clarified in Eq.~\eqref{initial_cond_model_1}. On the other hand, for Model II, we have followed the same approach, constraining all the necessary parameters except $\mathcal{M}_{0}$. The parameter $\mathcal{M}_{0}$ is fixed to a certain value \(\mathcal{M}_0 = 8.0\), as this parameter must be a positive number. We emphasize that although this parameter is fixed, the field variables \((x_0, y_0)\) are still varying, as can be seen from the Hubble equation Eq.~\eqref{constraint_eq_hub_expo}, which reads as:
			\begin{equation}
				y_0 = \frac{\log \left(-\frac{12 (\Omega_{m0}-1)}{\mathcal{M}_0 \left(3 x^4-2 x^2\right)}\right)}{\beta }.
			\end{equation}
		}
		{Our analysis shows that both models predict the value of \(H_0\) in close agreement with each other, but with a significant deviation from the \(\Lambda\)CDM model. The present-time matter energy density \(\Omega_{m0}\) is predicted to be lower for all datasets in both models compared to \(\Lambda\)CDM. 
			
			The evolution of cosmological parameters in Model I for the best-fit values provided in Tab. [\ref{tab:model1_confidence}], such as the fractional energy densities and the effective equation of state, is shown in Fig. [\ref{fig:comp_evo_kes_power}]. In this plot, the evolution is depicted against the number of e-folds \(N\). The results demonstrate that the model exhibits an extended matter-dominated phase (i.e., \(\Omega_{m} > \Omega_{\phi}\)), for all datasets, before transitioning to an accelerating epoch. During this matter-dominated phase, the effective equation of state is nearly zero, indicating attractive scaling behavior. As the system enters the late-time phase, the model shows phantom behavior for CC+BAO+DESBAO+PP+DES data, \(\omega_{\rm eff} \sim -1.05\) and accelerating behavior for rest of the datasets.
			
			\begin{figure}[h!]
				\centering
				\subfloat[\label{fig:lcdm_bestfit}]{\includegraphics[scale=0.7]{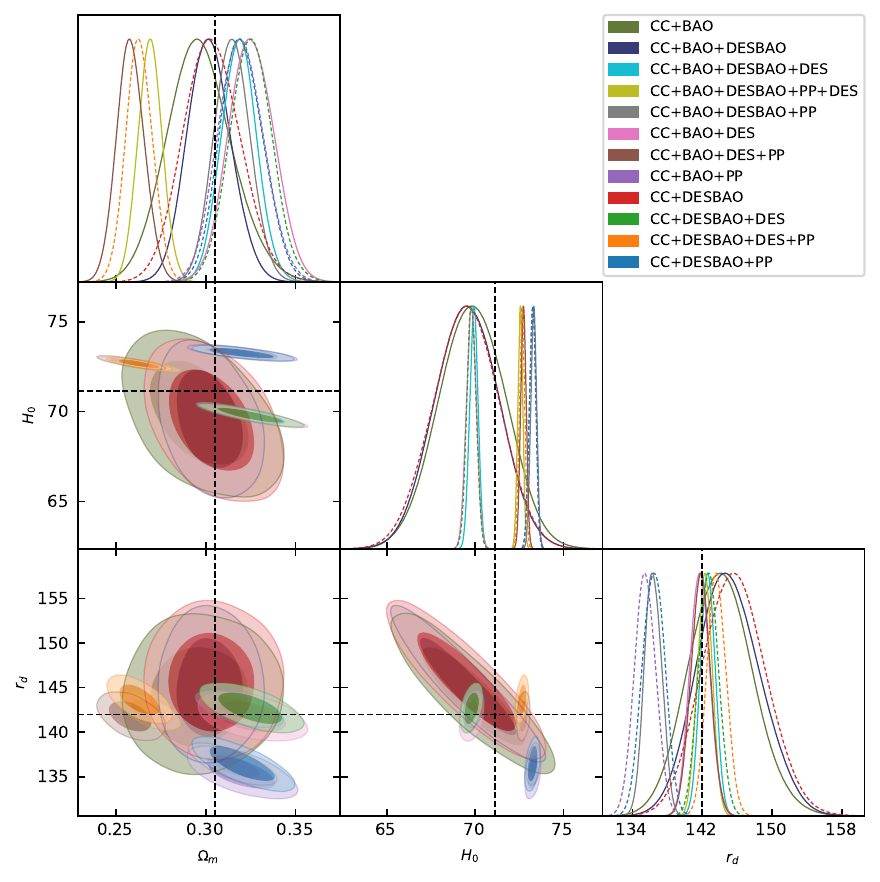}}\\
				\subfloat[\label{fig:kes_power_bestfit}]{	\includegraphics[scale=0.55]{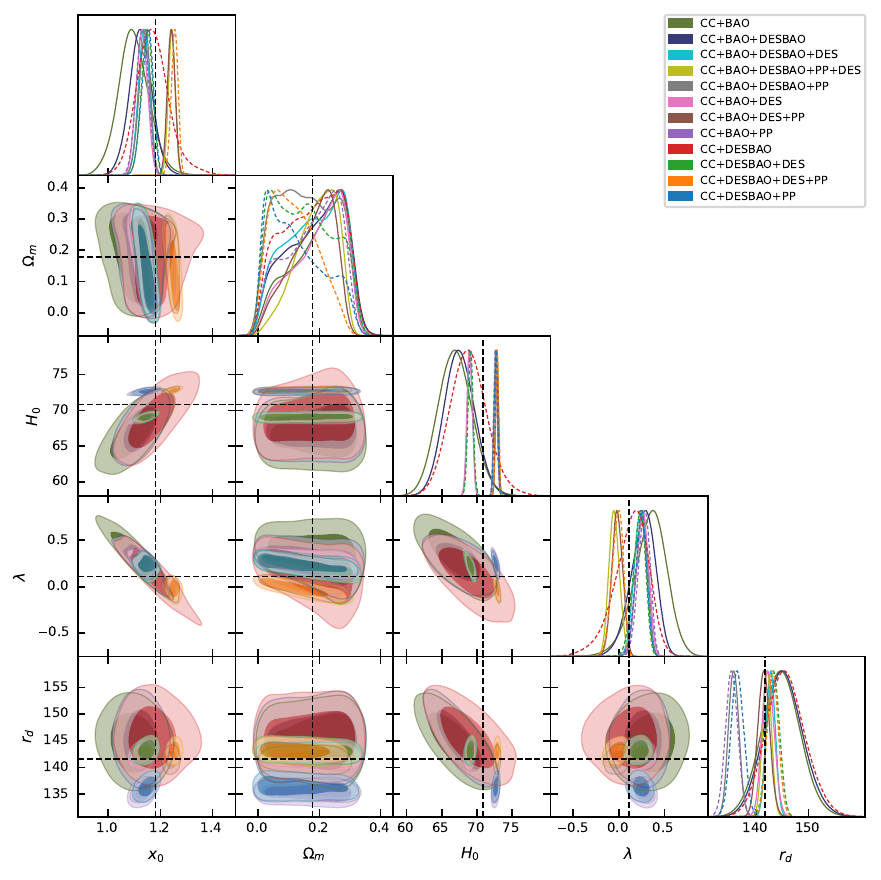}}
				\subfloat[\label{fig:kes_expo_bestfit}]{	\includegraphics[scale=0.55]{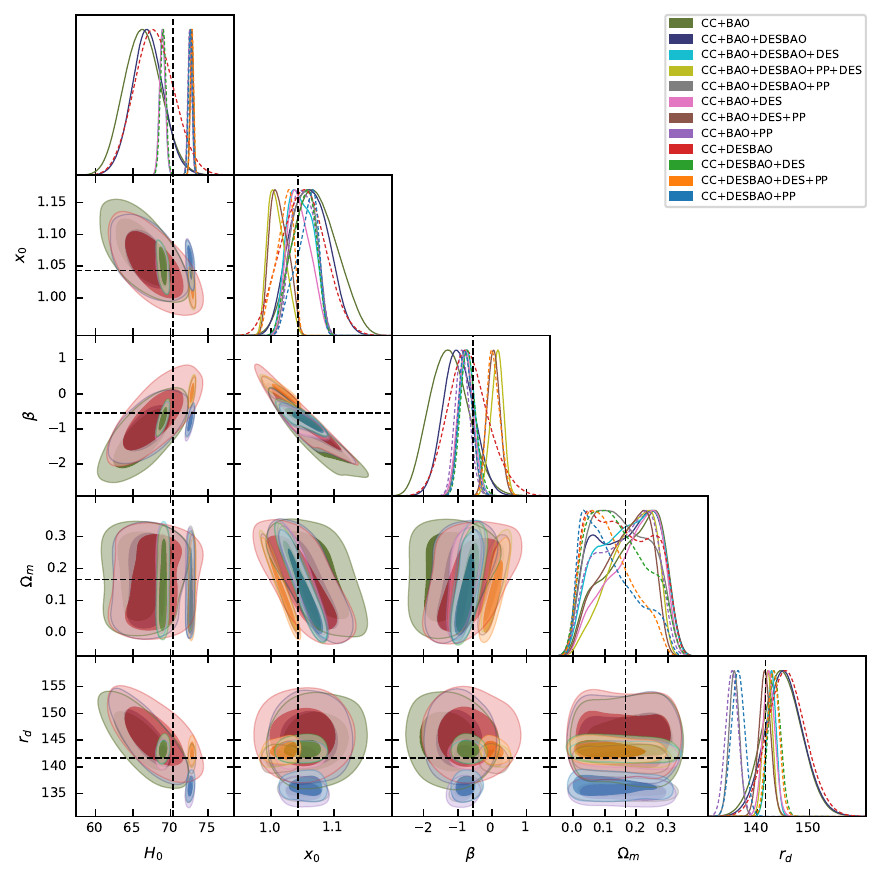}}
				
				\caption{1D and 2D marginalized posterior distribution of the model parameters of (a) the flat \(\Lambda\)CDM and the $k$-essence field with (b) inverse square type and (c) exponential type potential with fixed $\mathcal{M}_0 = 8$ for different datasets.}
				\label{fig:triangle_plot}
			\end{figure}

			The cosmological parameters of Model II, corresponding to the best-fit values, are plotted against \(\tilde{N}\) in Fig. [\ref{fig:evo_kes_exp_best}], as the autonomous system of equations is regulated at the coordinate \(z = -1\). In this case, a matter-dominated phase persists for an extended period during the past epoch in terms of e-folds \(\tilde{N}\). As the system enters the current epoch, the field density increases to 70\%, leading to acceleration. In the future epoch, an interesting scenario arises where, the model exhibits an accelerating solution, but the effective EoS begins to increase towards positive values, indicating a phase transition from an accelerating to a non-accelerating scenario. On the other hand, for the CC+BAO+DESBAO+PP+DES, the model exhibits stable phantom behavior $\omega_{\rm eff} \to -1$.

			{Additionally, we evaluated the information criteria, specifically AIC and BIC, for the models. For this comparison, \(\Lambda\)CDM was considered the base model. The model having lower value of AIC or BIC is preferred over the other. From Tab.~[\ref{tab:lcdm}], [\ref{tab:model1_confidence}] and [\ref{tab:mode2_confidence}], it is evident from the AIC criterion that Model I and Model II are preferred over the $\Lambda$CDM model for most data combinations. However, according to the BIC, the $\Lambda$CDM model is preferred over Model I and Model II for all data combinations, as BIC more effectively penalizes models with extra number of parameters.  }}\\

		Before we end this section we want to opine on the pure kinetic $k$-essence models, which are thought to be simpler to handle as there are no dynamic potential terms and where the Lagrangian density of the $k$-essence field is purely a function of $X$. In these models $V$ is a constant parameter. 
		Such a case is discussed in ref. \cite{Scherrer:2004au}, where the author assumes that the Lagrangian of the $k$-essence field has an extrema at some point $X=X_0$. It is seen that if one Taylor expands the Lagrangian at $X=X_0$ then one obtains a theory where the energy density of the $k$-essence field near $X_0$ can be represented as a sum of two separate terms. One of these terms can be interpreted to be coming from a dark energy like component and the other term scales with the scale-factor as $a^{-3}$, specifying a dark matter like component. Pure kinetic $k$-essence fields can in a certain way unify dark matter and dark energy sectors. The resultant theory has some parameters whose values require to be ascertained by observational data. We see that with our present sets of data it is practically impossible to give any bound on the parameters appearing in the theory. Although a $k$-essence sector with a non trivial potential can be tackled by our method and we can properly give bounds on various parameters and initial conditions appearing in the theory, it is pratically impossible to do such an analysis with a less complicated model. One of the reasons why this happens is perhaps related to the high level of generality of the problem discussed in  ref. \cite{Scherrer:2004au}. This theory is very general in nature and does not specify any particular form of the Lagrangian of the purely kinetic $k$-essence. Due to this general feature the energy density of the field near the extrema can be expressed in terms of six constants ($V_0$, $X_0$, $F_0$, $F_2$, $a_1$ and $\epsilon_1$). It becomes difficult to put reasonable bounds on all these six parameters using the present set of observational data. As we do not get any positive result for this case we do not present any more detail about this theory here. Readers who want to get a feeling of the difficulty one faces, if one wants to compare it with data, can consult Appendix \ref{app1} where the problem is briefly discussed.
		
		\section{Conclusion}
		\label{sec:conclusion}

		{
			In this study, we analyzed the \(k\)-essence scalar field model, particularly the Lagrangian form \(\mathcal{L}_{\phi} = - V(\phi) (-X + X^2)\), with an inverse square type potential (Model I) and an exponential potential (Model II), comparing their predictions with the \(\Lambda\)CDM model using observational probes consisting of CC, BAO, DESBAO, Pantheon+SH0ES and DES Supernova data sets. We studied the scalar field models by redefining their field equations in a dynamical system framework and obtained the necessary conditions on the model parameters to generate different phases of the universe, namely (Radiation \(\longrightarrow\) Matter \(\longrightarrow\) Accelerating \(\overset{\text{or}}{\longrightarrow}\) Phantom \(\longrightarrow\) Deceleration).
			
			Many studies have been conducted with the inverse square potential, but few have analyzed the exponential type potential with this particular form of the Lagrangian. One key challenge in the dynamics of the exponential potential is the need for more than two dynamical variables to close the autonomous system of equations, resulting in a 3D phase space. Due to the different structure of dynamical variables compared to the inverse square potential, the exponential model does not yield critical points that signify the current accelerating or phantom phase. The critical points only describe either far past or future epochs. Thus, model parameters such as \(\beta\) and \(\mathcal{M}_0\) could not be constrained with the arguments presented for Model I. Therefore, we relied on numerical evolution by varying the initial conditions and model parameters within a specific range to see whether the model produces different phases of the universe. We found that for certain ranges, the model produces a matter-dominated phase, and as the system enters the current epoch, the system's equation of state becomes lower than the background fluid's EoS, exhibiting tracking behavior. Thus, the exponential model can produce a stable accelerating (phantom) phase followed by the matter phase for certain ranges, alleviating the fine-tuning issue.
			
			However, the picture becomes clearer after employing the Markov Chain Monte Carlo (MCMC) method for parameter inference. The analysis demonstrated that the inverse square type model can produce a late-time accelerating scenario for all datasets. Notably, for certain datasets, the model exhibits a phantom behavior with \(\omega_{\rm eff} < -1\) in the future epoch, indicating a significant deviation from the concordance \(\Lambda\)CDM model.

			On the other hand, Model II, which corresponds to the exponential type potential, displays a similar scenario to Model I, where the matter phase dominates for a certain period, and the matter energy density exceeds the field energy density. However, an intriguing aspect of this model emerges: in the late-time phase, for the combined datasets, the model demonstrates a cosmological constant-type behavior, persisting over an extended period. Additionally, for the other datasets, the model undergoes a phase transition in the future epoch, shifting from an accelerating phase to a non-accelerating phase. This significant difference adds complexity to the model's dynamics and suggests that further studies, particularly with Planck datasets, are necessary to place stringent constraints on the model parameters.
			
			{We further assessed the viability of the models using statistical criteria, specifically the Akaike Information Criterion (AIC) and the Bayesian Information Criterion (BIC), and compared the results with those of the \(\Lambda\)CDM model. For most of the combined datasets, both models outperformed \(\Lambda\)CDM according to AIC criterion. However, \(\Lambda\)CDM remained more favorable for all the data combinations when we considered BIC criterion. This is because BIC criterion penalizes models with additional parameters. This suggests that while the current models provide good fit to the all the data combinations, the $\Lambda$CDM model is preferred as compared to $k$-essence models.}
			
		{However, recent studies indicate that we require a cosmological model that enhance the late-time cosmic acceleration to alleviate the Hubble tension. Phantom dark energy models, characterized by an equation of state parameter $w$ evolving just below $-1$, could potentially resolve the Hubble tension \cite{DiValentino:2020naf, DAHMANI2023101266, El-Zant:2018bsc, Bag:2021cqm}. The increase in acceleration within the phantom dark energy models raises the best-fit value of $H_0$ as compared to the $\Lambda$CDM model, brings it closer to the values derived from local distance ladder measurements. The $k$-essence models considered in this study showing phantom nature can potentially alleviate the Hubble tension. However, achieving an $H_0$ value from the CMB data that aligns with local measurements is insufficient on its own; a superior statistical fit of phantom models relative to the standard $\Lambda$CDM model and dynamical dark energy models is also required. Further analysis, including CMB data, is crucial to determine whether the $k$-essence models produce an $H_0$ value that is more consistent with the SH0ES measurement and to determine whether the $k$-essence models produce superior fit to CMB data as compared to the $\Lambda$CDM.}} \\
		     {Before we conclude, it is important to note that although the combined data sets marginally favored the phantom nature of the model, phantom phases of accelerated expansion are not physically well motivated due to their ability of producing a diverging phantom energy density in finite cosmological time \cite{Caldwell:2003vq}. Except this there can be other difficulties in phantom cosmology \cite{Ludwick:2017tox}.  In our analysis, for both models (corresponding to different potentials), the universe predominantly exhibits non-phantom behavior in the late-time phase, except for the dataset combination CC+BAO+DESBAO+PP+DES. This fact shows that our work is not unduly biased for phantom like behavior of the $k$-essence sector but as $k$-essence theories allow phantom like phases, our result to a certain extent, depends on phantom like behavior of the $k$-essence field. We can conclude by saying that alternative approaches beyond phantom dark energy models, along with additional data points at intermediate redshifts, are essential to effectively address the Hubble tension problem.}

		\begin{acknowledgments}
			One of the authors Sarath Nelleri is thankful to IIT Kanpur for providing with the institute postdoctoral fellowship. We would like to acknowledge the use of PARAM Sanganak, a high-performance computing facility at the Indian Institute of Technology Kanpur (IITK), for carrying out the data analysis work.
		\end{acknowledgments}
		
		\appendix
		
		\section{Critical Points at infinity in the case of power law potential\label{appen:cric_inf}}
		{
			In this appendix, we derive the critical points for the power-law potential by transforming the predefined variables. The ranges of the predefined variables are:  
			\begin{equation}
				-\infty < x < +\infty, \quad 0 < y < +\infty, \quad 0 \leq \Omega_{m} < 1\ .
			\end{equation}  
			Using these variables, the Hubble constraint is expressed as:  
			\begin{equation}
				y^2 = 1 - \Omega_{m} + x^2\ .
			\end{equation}  
			Since \(\Omega_{m}\) is restricted to a finite range, for very large \(x \to \infty\), it follows that $y$ also tends to infinity. To address this asymptotic behavior, where both \(x\) and \(y\) diverge, we introduce a transformation of variables using hyperbolic trigonometric functions:  
			\begin{equation}
				x = u \sinh(v), \quad y = u \cosh(v)\, .
			\end{equation}  
			In these new variables, the Hubble constraint becomes:  
			\begin{equation}
				1 = u^2 + \Omega_{m}\ .
			\end{equation}  
			Here, the fractional energy density of the field is given by \(\Omega_{\phi} = u^2\), which is constrained within the range \(0 \leq u^2 \leq 1\). Thus, the variable \(u\) is now restricted to a finite range.  
			
			The other variable, \(v\), is related to \(x\) and \(y\) as follows:  
			\begin{equation}
				v = \tanh^{-1}(x/y)\ .
			\end{equation}  
			The range of \(v\) remains:  
			\begin{equation}
				-\infty < v < +\infty\ .
			\end{equation}  
			To compactify \(v\) within a finite range, allowing for the identification of critical points at infinity, we apply the following Poincaré transformation:  
			\begin{equation}
				\bar{v} = \frac{v}{\sqrt{1 + v^2}}\ .
			\end{equation}  
			This maps \(\bar{v}\) into the range:  
			\begin{equation}
				-1 < \bar{v} < +1\ .
			\end{equation}  
			Using the transformed variables \((u, \bar{v})\), the dynamics of the system are now confined to finite ranges. The corresponding autonomous system of equations becomes: 
			\begin{eqnarray}
				u' & = & -\frac{1}{2} u \left(u^2-1\right) \left(\cosh \left(\frac{2 \bar{v}}{\sqrt{1-\bar{v}^2}}\right)-2\right),\\
				\bar{v}' & = & -\frac{\left(1-\bar{v}^2\right)^{3/2} \coth \left(\frac{\bar{v}}{\sqrt{1-\bar{v}^2}}\right) \left(\sqrt{6} \lambda  u \sinh \left(\frac{\bar{v}}{\sqrt{1-\bar{v}^2}}\right)+\cosh \left(\frac{2 \bar{v}}{\sqrt{1-\bar{v}^2}}\right)-5\right)}{4 \coth ^2\left(\frac{\bar{v}}{\sqrt{1-\bar{v}^2}}\right)-2}.
			\end{eqnarray}
			
			\begin{table}[t]
				\centering
				\begin{tabular}{ccccccc}
					\hline
					Points & \(u\) & \(\bar{v}\) & $\Omega_\phi$ & $\Omega_{m}$ &  $\omega_{\rm eff}$ & Stability \\
					\hline
					$P_{1}$ & 0 &0.753 & 0 & 1 & 0 & Saddle \\
					\hline
					$P_{2}$ & 1 & $\frac{\cosh ^{-1}\left(0.5 \sqrt{3. \lambda ^2-1. \sqrt{9. \lambda ^4+48. \lambda ^2}+12.}\right)}{\sqrt{\cosh ^{-1}\left(0.5 \sqrt{3. \lambda ^2-1. \sqrt{9. \lambda ^4+48. \lambda ^2}+12.}\right)^2+1.}}$ & 1 & 0 & Fig. [\ref{fig:eos_inf_p2_power}] & Fig. [\ref{fig:stab_inf_p2_power}] \\
					\hline
					$P_{3}$ & 1 & $\frac{\cosh ^{-1}\left(0.5 \sqrt{3. \lambda ^2+\sqrt{9. \lambda ^4+48. \lambda ^2}+12.}\right)}{\sqrt{\cosh ^{-1}\left(0.5 \sqrt{3. \lambda ^2+\sqrt{9. \lambda ^4+48. \lambda ^2}+12.}\right)^2+1.}}$ & 1 & 0 & Fig. [\ref{fig:eos_inf_p3_power}] & Fig. [\ref{fig:stab_inf_p3_power}]\\
					\hline
					$P_{4}$ & 1 & 0 & 1 & 0 & 1/3 & Unstable\\
					\hline
					$P_{5}$ & 0 & 1 & 0& 1& 0 & Unstable\\
					\hline 
					$P_{6}$ & $\frac{\sqrt{3}}{\lambda}$ & $0.55$ & $3/\lambda^2$ & $1-\frac{3.}{\lambda ^2}$ & 0 & Stable ($\lambda>\sqrt{3}$)\\
					\hline
					\hline
				\end{tabular}
				\caption{Critical points for the compactified variables \((u, \bar{v})\) in the context of the power-law potential.}
				\label{tab:newcric_power}
			\end{table}
			
			Here, the prime denotes differentiation with respect to the logarithmic time variable \(dN = d \log(a)\). The autonomous system reveals invariant sub-manifolds at  \(u = 0\) and \(v = 0\) lines. These invariant manifolds ensure that trajectories originating in the \(u > 0, \bar{v} > 0\) quadrant remain confined within it.
			
			The system yields six critical points, which are summarized in Table [\ref{tab:newcric_power}]. These points encapsulate the evolutionary sequence of the universe: radiation \(\longrightarrow\) matter \(\longrightarrow\) accelerating \(\overset{\text{or}}{\longrightarrow}\) phantom phases. Below, we briefly describe these critical points, focusing on their fractional energy densities and effective equation of state (EoS).
			
			\begin{itemize}
				\item \textbf{Points $P_{1,5}$:} Both points correspond to fluid-dominated energy density with a vanishing effective EoS, indicative of a matter-dominated phase. However, \(P_1\) exhibits a saddle nature, while \(P_5\) is unstable. 
				
				\item \textbf{Point $P_{4}$:} This point is characterized by a field-dominated energy density and an effective EoS of \(1/3\), representing non-thermal radiation. This point is unstable.
				
				\item \textbf{Points $P_{2,3}$:} These points depend on the parameter \(\lambda\) and correspond to field-dominated phases. \(P_2\) leads to an accelerating solution, while \(P_3\) results in a phantom phase. Figures [\ref{fig:eos_inf_p2_power}] and [\ref{fig:eos_inf_p3_power}] depict the variation of the effective EoS with \(\lambda\), while Figures [\ref{fig:stab_inf_p2_power}] and [\ref{fig:stab_inf_p3_power}] show the eigenvalue analysis, highlighting their stability. 
				
				\item \textbf{Point $P_{6}$:} This point exists only for \(\lambda \geq \sqrt{3}\). While it represents a stable matter-dominated phase, it is not physically viable, as the universe cannot remain stabilized in the matter-dominated epoch.  
			\end{itemize}
			
			Overall, this analysis reveals a dynamical behavior similar to the original variables, with the additional presence of critical points at infinity. However, these points do not affect the late-time dynamics since they are non-accelerating. 
			
			The overall behavior of the system becomes evident from the phase-space plot in Figure [\ref{fig:phase_inf_power_law}] for different values of \(\lambda\). For specific \(\lambda\), the point \(P_6\) violates the Hubble constraint, rendering it invisible in the phase space. The gray-shaded region represents the allowed domain defined by \(0 \leq \Omega_{\phi} < 1\). The phase space is divided into regions with distinct dynamics: The green region corresponds to the accelerating regime (\(-1 \leq \omega_{\rm eff} \leq -1/3\)). The red region indicates the phantom regime (\(\omega_{\rm eff} < -1\)). The yellow region represents the radiation-dominated phase (\(0.29 < \omega_{\rm eff} < 0.31\)). The phase-space trajectories originating from \(P_4\) initially converge towards \(P_1\), corresponding to the matter-dominated phase. Since \(P_1\) is a saddle point, these trajectories are repelled and eventually stabilize at the accelerating attractor \(P_2\) (for \(\lambda = 0.2\)). Similarly, trajectories from \(P_5\) also stabilize at \(P_2\). However, a few trajectories bypass move directly toward the stable phantom point \(P_3\).
			
			For \(\lambda = -0.2\), the trajectories originating from \(P_4\) also stabilize at the phantom point \(P_3\). For \(\lambda > 1\), the trajectories stabilize at \(P_2\) or \(P_3\), but neither of these points lies in the accelerating or phantom regions, making the dynamics irrelevant for this case. } 
		
		{It is seen that only point $P_4$ has one of the coordinates $\bar{v}$ to be 0 designating that it corresponds to the point at where both $x$ and $y$ tend to $\infty$. We did not discuss about this critical point in our main analysis given in the text as we were mainly interested in the critical points in the finite regions of phase space.   For $\bar{v}=0$ we must have both $x$ and $y$ tends to infinity (but $y$ tends to infinity at a much faster rate than $x$) such that $x/y$ tends to zero. This point does not play a prominent role in the late time universe and consequently only the critical points in the finite region of phase space remain relevant for our purpose \footnote{One must note that the point with $\bar{v}=1$ does not specify any of our old variables $x$ or $y$ are infinity. Although this point designates $v \to \infty$, but $x$ and $y$ are finite for this value of $v$.} } {In summary, the compactified phase space provides a richer dynamical perspective compared to the original variables. Nevertheless, the overall behavior remains consistent, except for the existence of a critical point at infinity in the compactified variables. Importantly, these points do not influence the dynamics at the current epoch, justifying the use of the original variables for the data analysis presented in Sec. \ref{sec:data_analysis}.

			\begin{figure}
				\centering
				\subfloat[\label{fig:eos_inf_p2_power}]{\includegraphics[scale=0.45]{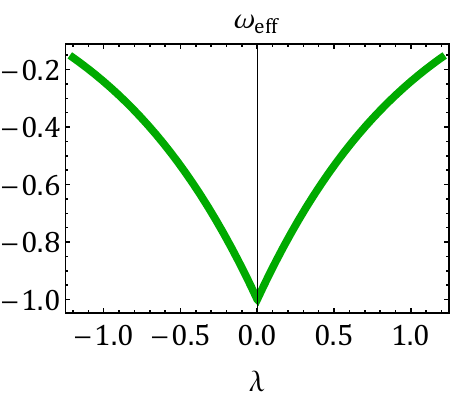}}
				\hspace{0.2cm}
				\subfloat[\label{fig:stab_inf_p2_power}]{\includegraphics[scale=0.45]{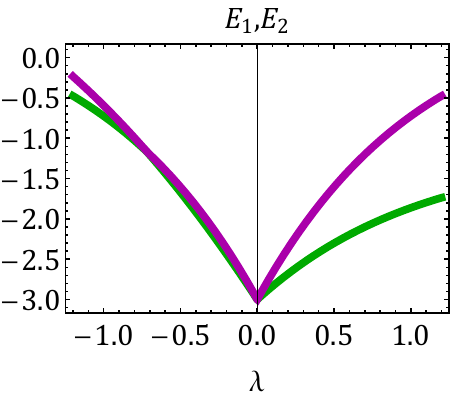}}
				\hspace{0.2cm}
				\subfloat[\label{fig:eos_inf_p3_power}]{\includegraphics[scale=0.45]{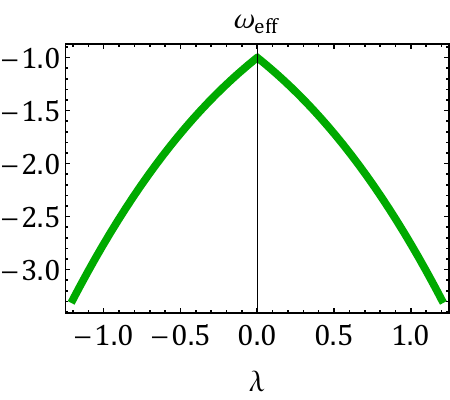}}
				\hspace{0.1cm}
				\subfloat[\label{fig:stab_inf_p3_power}]{\includegraphics[scale=0.45]{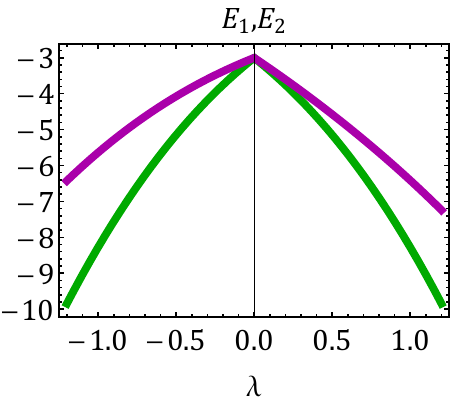}}
				\caption{Variation of the effective equation of state (EoS) and eigenvalues for the power-law potential in compactified variables \((u, \bar{v})\): (a) and (b) correspond to point \(P_{2}\), while (c) and (d) correspond to point \(P_{3}\).}
			\end{figure}

			\begin{figure}
				\centering
				\includegraphics[scale=0.6]{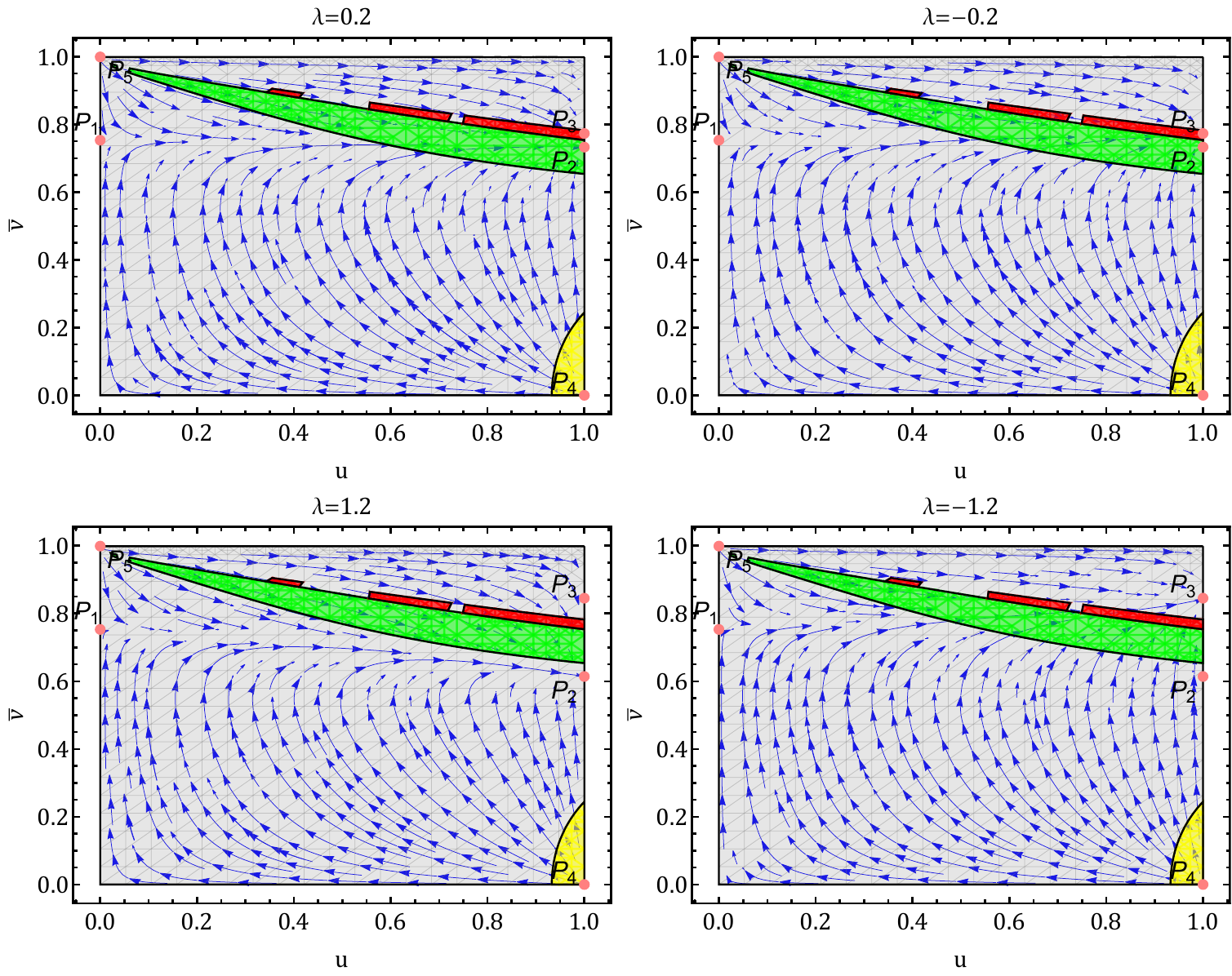}
				\caption{Phase space for the power-law potential using the compactified variables \((u, \bar{v})\), shown for different values of $\lambda$.}
				\label{fig:phase_inf_power_law}
			\end{figure}
		}
		
		\section{A compact phase space analysis for the case of exponential potential\label{appen:2}}
		{
			In this section, we discuss the compactification of dynamical variables for an exponential type potential. The predefined ranges for these variables are:
			\begin{equation}
				- \infty < x< + \infty, \quad  	- \infty < y< + \infty, \quad  -1 < z < 1.
			\end{equation}
			To compactify \( y \), we apply the Poincaré transformation, defining a new variable \( \bar{y} \) as:
			\begin{equation}
				\bar{y} = \frac{y}{\sqrt{1 + y^2}},
			\end{equation}
			with this transformation, \( \bar{y} \) is now constrained within:
			\begin{equation}
				-1 < \bar{y} < +1. 
			\end{equation}
			In these transformed variables, the Hubble constraint equation becomes:  
			\begin{equation}  
				1 = \Omega_{m} + \frac{\mathcal{M}_0 \left(\frac{3 x^4}{4} - \frac{x^2}{2}\right) (1-z)^2 e^{\frac{\beta  \bar{y}}{\sqrt{1-\bar{y}^2}}}}{3 (z+1)^2}.  
			\end{equation}  
			It is evident that by compactifying \( y \) into \(\bar{y}\), the variable \( x \) is also compactified. The autonomous system of equations now becomes:
			\begin{eqnarray}
				x' & =& \frac{-3 x^3+\frac{\beta  \left(\frac{x^2}{2}-\frac{3 x^4}{4}\right) (1-z)}{z+1}+3 x}{3 x^2-1}, \\
				\bar{y}' & =& \frac{x (1-z)}{z+1} \left(1-\bar{y}^2\right)^{3/2}, \label{yb_prime_expo}\\
				z' & =& -\frac{3 ((1-z) (z+1)) \left(\frac{1}{3} \mathcal{M}_0 \left(\frac{x^4}{4}-\frac{x^2}{2}\right) \left(\frac{1-z}{z+1}\right)^2 \exp \left(\beta  \frac{\bar{y}}{\sqrt{1-\bar{y}^2}}\right)+1\right)}{4}\, .
			\end{eqnarray}
			With this transformation, we encounter a similar issue (as faced previously when dealing with the equations in the main text): the autonomous system diverges as \( z \to -1 \). To address this, we redefine the time variable as \( dN \to (1 + z) d\tilde{N} \). The resulting equations can be written down in a straightforward manner although we have not written them here.
			A potential concern is the behavior of the exponential argument in \( z' \) as \( \bar{y} \to 1 \), which might cause \( z' \) to diverge. However, the structure of \( z' \) includes the factor \( (1 - z)(1 + z) \), so when \( \bar{y} \) approaches 1, \( z \) simultaneously approaches either \( \pm 1 \), keeping \( z' \) finite. Thus, the autonomous system remains differentiable, and \( \bar{y} \to \pm 1 \) (or equivalently, \( y \to \mp \infty \)) remains a valid limit. This behavior is illustrated in Fig. [\ref{fig:evo_kes_entire_expo}], for the non-transformed variable evolution.
			
			In this case, the critical points listed in Tab. [\ref{tab:cric_expo_poincare}] are identical to those for the non-transformed variables (as presented in the main text), so further discussion on them is unnecessary.} 
		\begin{table}[t]
			\centering
			\begin{tabular}{ccccc}
				\hline
				Points & $(x,\bar{y},z)$ & $\Omega_{\phi}$ & $\Omega_{m}$ & $\omega_{\rm eff}$ \\
				\hline
				$P_{1,2}$ & $(0, \text{Any}, \mp 1)$ & 0 & 1 & 0 \\

				\hline
				$P_{3,4}$ & $(\mp 1, \text{Any}, 1)$ & 0 & 1 & 0\\
				
				\hline
				\hline
			\end{tabular}
			\caption{The nature of critical points corresponding to an exponential potential with compacted $\bar{y}$.}
			\label{tab:cric_expo_poincare}
		\end{table}
		{ It can be seen that for all the critical points in this case ($H\to \infty$ or $H\to 0$), the effective EoS  of the system is zero and consequently these critical points are not important for producing accelerated expansion of the universe.  Moreover as far as comparison with data is concerned we get important results, which matches with our expectations about the present day universe, showing that a probable future or past asymptotic fixed point with unnatural values of the Hubble parameter, does not affect our predictions. 
			In such a case a critical point analysis does not serve our purpose.}
		
\begin{figure}[t]
			\centering
			\includegraphics[scale=0.8]{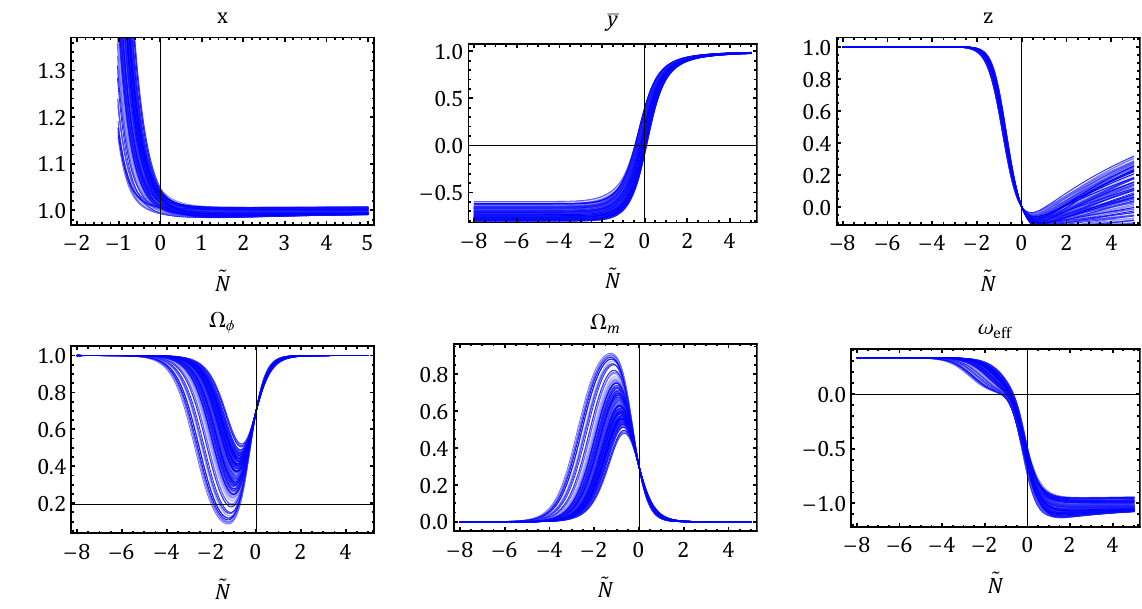}
			\caption{Numerical evolution of the system with an exponential potential, using the compactified variable \(\bar{y}\).}
			\label{fig:evo_expo_compact}
		\end{figure}

		{In our case we work out the dynamics using numerical simulations varying initial conditions and model parameters within the ranges: 
			\begin{equation}
				-0.1 < \bar{y}_0< 0.4, \quad z_0  = 0, \quad 6 <\mathcal{M}_0 <8, \quad -0.1 <\beta < 0.4, \quad \Omega_{\rm m0} = 0.3\,.
			\end{equation}
			The initial condition for \( x \) can be determined from the constraint relation. The numerical evolution is shown in Fig. [\ref{fig:evo_expo_compact}]. In the past epoch, \( x \) takes larger values, but the fractional energy densities for both the field and fluid remain finite and within physically acceptable limits. In the future epoch, \( \bar{y} \) approaches \( 1 \), where field energy density dominates, and matter density vanishes, signifying a dark energy-dominated behavior with an effective EoS transitioning between accelerating and phantom regimes. This indicates that a \( k \)-essence field with an exponential potential can produce stable accelerating or phantom solutions at late times, with behavior similar to that shown in Fig. [\ref{fig:evo_kes_entire_expo}]. 
		}
		
		\section{A note on pure kinetic $k$-essence }
		\label{app1}
		\color{black}{
			In ref. \cite{Scherrer:2004au}, the author demonstrated that the $k$-essence field with a constant potential can unify the dark matter and dark energy phases. In this case, the pure $k$-essence Lagrangian becomes:
			\begin{equation}
				P(\phi, X) = V_0 F(X),
			\end{equation}
			where \(P(\phi, X)\) denotes the pressure of the $k$-essence field and \(V_0\) is the constant potential. In the present case we do not assume any specific form of $F(X)$, the only relevant assumption is that $F(X)$ has an extremum at $X=X_0$. In the aforementioned reference it was shown that if such an extremum exists it will be stable under small perturbations around $X_0$. In ref. \cite{Scherrer:2004au} it was shown that the Taylor series expansion of \(F(X)\) near its extremum \(X_0\) becomes:
			\begin{equation}
				F(X) = F_0 + F_2 (X - X_0)^2,
			\end{equation}
			where \(F_0\) and \(F_2\) are constants. The corresponding energy density becomes:
			\begin{equation}
				\rho = V_0 \left(-F_0 + 4 F_2 X_0^2 \epsilon_1 (a/a_1)^{-3}\right),
			\end{equation}
			where \(a_1\) is an integration constant with \(\epsilon_1 \ll 1\). To constrain the model parameters against the data, we express the Friedmann equation as:
			\begin{equation}
				3H^2 = \kappa^2 V_0 \left(-F_0 + 4 F_2 X_0^2 \epsilon_1 (a/a_1)^{-3}\right).
			\end{equation}
			Rewriting this equation in terms of the redshift \(z\):
			\begin{equation}
				H(z) = H_0 \sqrt{\Omega_{F_0} - 4 \frac{F_2}{F_0} \Omega_{F_0} X_0^2 \epsilon_1 a_1^{3} (1+z)^3},
			\end{equation}
			where \(\Omega_{F_0} = \frac{-F_0 \kappa^2 V_0}{3 H_0^2}\). We can express other constants in terms of \(\Omega_{F_0}\) by using the condition that at \(z=0\), \(H=H_0\):
			\begin{equation}
				\epsilon_1 = \frac{F_0}{4 F_2 X_0^2 a_1^3} \left(1 - \frac{1}{\Omega_{F_0}}\right).
			\end{equation}
			Inserting this into the above equation results in:
			\begin{equation}
				H(z) = H_0 \sqrt{\Omega_{F_0} - \Omega_{F_0}(1 - 1/\Omega_{F_0}) (1+z)^3},
			\end{equation}
			which takes the same form as \(\Lambda\)CDM. From this, we see that it is not possible to constrain all the model parameters of the kinetic k-essence. One can only obtain constraints on \(\Omega_{F_0}\), and from this information, one can at best estimate the range of some parameters by assuming certain values for other parameters. Here, we will get the same value for $\Omega_{F_0}$ as $\Omega_{\Lambda0}$ as determined in Tab. [\ref{tab:lcdm}].

			\bibliographystyle{JHEP}
			
			\bibliography{ref}
			
		\end{document}